\documentclass[a4paper,fleqn,usenatbib]{mnras}
\usepackage{subfigure}
\usepackage{rotating}
\usepackage{newtxtext,newtxmath}
\usepackage[T1]{fontenc}
\usepackage{ae,aecompl}
\usepackage{tikz}
\usepackage{amsmath}
\usepackage{nccmath}
\usepackage{multicol, blindtext}
\usepackage{graphicx}	% Including figure files
\usepackage{amsmath}	% Advanced maths commands
\usepackage{float}		% Bind figures and plots to sections 
\usepackage{placeins}
\usepackage{array}
\usepackage{epsfig}

\newcommand{\Ms}{M$_{\odot}$}
\newcommand{\Ls}{L$_{\odot}$}

\title[Supermassive Stars in MESA]{Modeling Supermassive Primordial Stars with MESA}
\author[Herrington et al.]{Nicholas P. Herrington,$^{1}$\thanks{E-mail: nh478@exeter.ac.uk} 
Daniel J. Whalen,$^2$
Tyrone E. Woods$^{3}$
\\
\\
$^1$School of Physics and Astronomy, University of Exeter, Stocker Road, Exeter, EX4 4QL, UK \\
$^2$Institute of Cosmology and Gravitation, University of Portsmouth, Dennis Sciama Building, Portsmouth PO1 3FX, UK \\
$^3$National Research Council of Canada, Herzberg Astronomy \& Astrophysics Research Centre, 5071 West Saanich Road, \\
Victoria, BC V9E 2E7, Canada \\
}

\date{Accepted XXX. Received YYY; in original form ZZZ}

\begin{document}
\pagerange{\pageref{firstpage}--\pageref{lastpage}}
\maketitle

\begin{abstract}

Supermassive stars forming at $z \sim$ 15 - 20 are one of the leading contenders for the origin of the first quasars, over 200 of which have now been discovered at $z >$ 6. These stars likely form in pristine, atomically cooled haloes immersed in strong Lyman-Werner UV backgrounds or in highly supersonic baryon streaming flows. Atomic cooling triggers catastrophic baryon collapse capable of building up stars at rates of up to $\sim$1 \Ms\ yr$^{-1}$. Here we examine the evolution of supermassive stars with a much larger and finer grid of accretion rates than in previous studies with the \texttt{MESA} stellar evolution code.  We find that their final masses range from 3.5 $\times$ 10$^3$ \Ms\ - 3.7 $\times$ 10$^5$ \Ms\ at accretion rates of 0.001 \Ms\ yr$^{-1}$ - 1 \Ms\ yr$^{-1}$, respectively.  We also find that supermassive star evolution diverges at accretion rates of 0.01 \Ms\ yr$^{-1}$ - 0.02 \Ms\ yr$^{-1}$, above which they evolve as cool red hypergiants along the Hayashi track and collapse via the general relativistic instability during central hydrogen burning, and below which they evolve as hot blue supergiants and collapse at the end of their nuclear burning lifetimes after exiting the main sequence.
    
\end{abstract}


\section{Introduction}

Supermassive stars (SMSs) are one of the leading candidates for the origin of the first quasars, more than 200 of which have been discovered at $z >$ 6 \citep{fan03,fan06}, including eight at $z >$ 7 \citep{mort11,wu15,ban18,smidt18,mats19,zhu20,wang21}. Until recently, they were thought to form at $z \sim$ 10 - 20 in primordial halos immersed in high Lyman-Werner UV backgrounds \citep{latif14,agarw15,anna15,anna17} or in highly supersonic baryon streaming motions \citep[BSMs;][]{lns14,hir17,srg17} that suppress normal star formation until they reach masses of 10$^7$ \Ms\ and virial temperatures of 10$^4$ K. These temperatures trigger atomic cooling that causes gas to collapse to the center of the halo at rates of up to $\sim$ 1 \Ms\ yr$^{-1}$ \citep{bl03,ln06,rh09b,latif15b}. Such flows create massive, hot accretion disks that build up either a single SMS or binaries and small multiples \citep{wise19, ret20,latif20a,pat21b,pat21a}.  However, it has now been found that the rare haloes capable of forming quasars by $z \sim$ 6 can form SMSs without the need for UV backgrounds, BSMs, or even atomic cooling \citep{latif22b}

SMSs have been the subject of analytical studies since the 1960s \citep[e.g.,][]{iben63,chandra64,fowler64,fowler66} and numerical simulations since the 1970s \citep[e.g.,][]{af72a,st79,fuller86,baum99,sun18,but18}, but it has only been recently that they have been studied in the extreme flows in which they form \citep{hos13,um16,tyr17,hle18b,tyr22a,tyr21a}. These models indicate that rapidly accreting SMSs can reach masses of a few 10$^5$ \Ms\, and lifetimes of 1 - 2 Myr before collapsing to direct-collapse black holes (DCBHs). SMSs are the leading contenders for the seeds of the first quasars because it is difficult for ordinary Population III (Pop III) star BHs to grow rapidly after birth because they form in low densities and in some cases are ejected from their host halos \citep{wan04,ket04,awa09,wf12,srd18}. DCBHs are born with much larger masses and in much higher densities in halos that retain their fuel supply, even when it is heated by X-rays \citep{jet13}.

Previous studies have mostly found that rapidly-accreting primordial stars evolve as cool, red hypergiants along the Hayashi limit, with surface temperatures of 5,000 - 10,000 K due to H$^-$ opacity in their atmospheres, at least until they reach $\sim$ 10$^5$ \Ms\ \citep{hos13}. \citet{hle18b} found that SMSs can remain cool even above these masses, reaching luminosities greater than 10$^{10}$ \Ls. In other studies, SMSs evolving from similar initial conditions soon settle onto hotter and bluer tracks with temperatures of 20,000 - 40,000 K \citep{tyr17}. \citet{hle18b} found that stars growing at low rates ($\lesssim$ 0.005 \Ms\ yr$^{-1}$) also evolve along blue tracks, as can stars with clumpy accretion due to fragmentation or turbulence in the disk (\citealt{sak15} -- but see \citealt{sak16b}). It is not clear if these differences arise from the opacities used in the models, code physics (such as the numerical treatment of convection), or resolution (see also \citealt{Soker2020} for additional discussion of convergence issues and \citealt{titans} for further discussion of supermassive star models).

Recent work suggests the $\Omega \Gamma$ limit, in which radiation pressure and centrifugal forces equal gravity in the star, restricted SMS rotational velocities to at most a few percent of Keplerian \citep{hle18a}.  Consequently, the flows that created such stars in the early Universe must have had efficient mechanisms for shedding angular momentum.  One possibility is the rise of 'bars within bars' instabilities on small scales \citep[e.g.,][]{wta08}.  Another is the rise of magnetic fields in the accretion disk of the star, first on small scales due to turbulent subgrid dynamos \citep{schob12} and then their subsequent amplification via the $\alpha \Gamma$ instability on larger scales \citep{ls15}.

The relatively low surface temperatures of SMSs in most models to date imply that radiation from the star cannot overcome the enormous ram pressures of catastrophic infall and that they form, evolve and die in the accretion flows that create them. This has been corroborated at the initial stages of SMS formation by radiation hydrodynamical simulations by \citet{ard18} and \citet{luo18} \citep[see also][]{aaron17}. Any ionizing UV from the star would also only heat the gas to at most a few times the temperatures at which it already falls onto the star without forming the usual hot, high-pressure bubbles that drive all the baryons out of less massive Pop III star-forming halos \citep{suh09,susa13,hir13,hir15,sug20,latif21a,latif22a}. 

Pop III SMSs accreting at rates of 0.1 - 1 \Ms\ yr$^{-1}$ would have extremely large luminosities that could, in spite of their low surface temperatures, be detected in the near infrared (NIR) by the {\em James Webb Space Telescope} ({\em JWST}) and large ground-based telescopes in the coming decade \citep{jlj12a,hos13,sur18a,sur19a,wet20a}.  Their BHs could also be found in the NIR at $z \sim 15 - 20$ by {\em JWST} \citep{pac15,nat17,bar18,wet20b} and at $z \sim$ 6 - 8 by {\em Euclid} and the {\em Roman Space Telescope}, although lensing by massive galaxies and galaxy clusters in their wide fields could extend these detections up to $z \sim$ 15 \citep{vik22a}. DCBHs will only be marginally visible to the Square Kilometre Array or next-generation Very Large Array at $z \gtrsim$ 6 - 8 \citep{wet20a,wet21a} but would become more luminous after growing to larger masses at later times \citep{wet22b}.  They could also be detected out to $z \sim$ 10 by future X-ray missions such as the {\em Advanced Telescope for High-Energy Astrophysics} ({\em ATHENA}) and {\em Lynx} \citep{athena,lynx}.

Previous studies of supermassive primordial star evolution considered small representative samples of 4 - 7 accretion rates in the range of 0.001 \Ms\ yr$^{-1}$ - 10 \Ms\ yr$^{-1}$. Here we revisit the evolution of supermassive primordial stars with the Modules for Experiments in Stellar Astrophysics (\texttt{MESA}) code by examining a much larger and finer grid of accretion rates than previously explored. We describe the physics in our runs and their setups in Section~2. The evolution of the stars is examined in detail in Section~3. We tally final masses for SMSs as a function of accretion rate in Section~4 and conclude in Section~5.

\section{Numerical Method}

\label{section:method}

\texttt{MESA} is a one-dimensional (1D) Lagrangian stellar evolution code that solves  equations of stellar structure that are implicitly coupled to convective mixing and nuclear burning \citep[version 12778;][]{paxt11,paxt13,paxt18}. \texttt{MESA} has implicit hydrodynamics that can solve for the velocity of each zone in the model at the onset of collapse.  We use the 21-isotope APPROX network, which  includes the pp chain, triple-alpha burning, and the CNO cycle. Our models have an equation of state (EOS) that is a composite of several datasets such as the OPAL/SCVH tables \citep{rogers02,saumon1995}, which are used at lower densities and temperatures like those in the outer regions of the star and its atmosphere, and the \texttt{HELM} and \texttt{PC} EOSs \citep{ts00,potekhin10}, which are applied to high densities and temperatures like those in the core of the star \citep[see Figure 1 and Section 4.2 of][]{paxt11}.  We use the Henyey method for convective mixing length theory, with the mixing length parameter $\alpha_{\mathrm{MLT}} =$ 2, and an overshooting parameter set to 0.1.  These two values are the defaults for massive stars in MESA and are consistent with those used in previous SMS simulations \citep[e.g.,][]{tyr17}.  We also include superadiabatic convection in radiative regions when the default criteria for it in MESA are met.  We also use the Ledoux criterion for convection along with the \texttt{MESA} prescription for smoothing composition gradients in non-convective regions that trail retreating convection zones.  This measure reduces unphysical steps in entropy that can develop in the interior of the star over time.  

The effects of general relativity (GR) on the structure of the star are approximated by the Tolman-Oppenheimer-Volkoff (TOV) correction to the equation of hydrostatic equilibrium,
\begin{equation}
       \frac{dP}{dr} = -\frac{Gm\rho}{r^2},
       \label{eqn:HE}
\end{equation}
which is implemented as a post-Newtonian correction to the gravitational constant, $G$,  
\begin{equation}
	G_{\textrm{rel}} = G\left(1 + \frac{P}{\rho c^2} + \frac{4\pi Pr^3}{m_r c^2} \right)\left(1 - \frac{2Gm_r}{rc^2}\right)^{-1}. 
	\label{eqn:grel}
\end{equation}
$G_{\textrm{rel}}$ is computed in every zone of the star every time step in the user-defined \textit{run\_star\_extras} subroutine in the \texttt{MESA}\_{\em star} subroutine and then used to update Equation~\ref{eqn:HE}.  We turn on post-Newtonian corrections to $G$ at the beginning of each run.  

\texttt{MESA} adaptively rezones the stars as they evolve.  Mesh refinement is triggered when gradients in temperature, pressure, and $^4$He abundances between adjacent zones exceed preset values: $\delta \log P/P >$ 1/30, $\delta \log T/T >$ 1/80, and $\delta \log(\chi + 0.01) >$ 1/20 $\log \chi$, where $\chi$ is the $^4$He mass fraction \citep[see section 6.4 of][]{paxt11}.  These criteria result in a larger number of zones usually being allocated near the center of the star to resolve nuclear burning and convective processes, and the stars are typically partitioned into about 1400 mass zones.  Accretion flows onto the star have a primordial composition that does not change as the star evolves.  Their entropy is matched to that of the surface of the star so it is assumed that the accretion luminosity is radiated away rather than deposited at the surface.  We exclude mass loss due to stellar winds because they are thought to be negligible at primordial compositions \citep{vink01,bhw01}.  We initialise all the stars as 20 \Ms\ fully-convective, $n =$ 1.5 polytropes with temperatures below 10$^6$ K to preclude nuclear burning \citep[Section 6.1 of][]{paxt11}.  They have primordial compositions, with mass fractions $\chi_{\mathrm{H}} =$  0.7516 and $\chi_{\mathrm{He}} =$ 0.2484.  Our 16 models are uniformly partitioned in log accretion rate over 3 decades: $10^{-3}$ \Ms\ yr$^{-1}$, $10^{-2}$ \Ms\ yr$^{-1}$ and $10^{-1}$ \Ms\ yr$^{-1}$.  

Models with accretion rates at or above 0.02 \Ms\ yr$^{-1}$ are first evolved up to 10,000 \Ms\ without hydrodynamics because it can lead to numerical instabilities if the core is still contracting or burning is about to begin.  Each model is then branched into two parallel runs:  with and without hydrodynamics.  Stars evolved with hydrodynamics eventually collapse because of pulsations induced by the GRI.  Stars without hydrodynamics cannot actually collapse because there are no fluid motions so it is instead inferred from the 'softening' of the EOS in the core of the star, when the adiabatic exponent falls below the critical value for radiation-dominated stars when GR is taken into account:
\begin{equation}
\Gamma < \frac{4}{3} + \frac{\beta}{6},
\end{equation}
where
\begin{equation}
\beta \sim  \frac{4 k_{\mathrm B}}{\mu s},
\end{equation}
$\mu$ is the mean molecular weight, and $s / k_{\mathrm B}$ is the entropy per baryon \citep{tyr17}.  We run both cases for each star to determine if the use of the softening of the EOS as a proxy for collapse produced accurate final stellar masses in previous studies.  Stars with hydrodynamics are evolved up to the mass at which their cores collapse irreversibly at velocities of 500 - 1000 km s$^{-1}$ and those without hydrodynamics are evolved until $\Gamma$ falls below the critical value in the core.  GR is turned on in these runs at all times.

Stars with accretion rates below 0.02 \Ms\ yr$^{-1}$ never reach masses at which they encounter post-Newtonian instabilities so GR is turned off in these models, but hydrodynamics is turned on from the beginning of the run in zones whose temperatures exceed 10$^7$ K.  Numerical instabilities that arise during the contraction of the core and the expansion of the star when it becomes fully convective after the depletion of central hydrogen make it difficult to evolve the stars to more advanced stages of burning so we halt their evolution at central H fractions of 1\% (He burning due to the triple $\alpha$ process begins well before this point).  However, our main objective is to determine the final mass of the star, and at these low accretion rates the star cannot gain much mass in the short times over which late stages of burning proceed.  We get to within a few percent of the true final mass of the star even though we exclude these stages.  We also activated a control that adds more resolution to the models when log $T_c > 8.15$, by which point the central hydrogen fraction in most of the stars has fallen to below 30\%.

Pressure must be imposed on the outer boundaries of these stars to prevent superluminal velocities from arising in their low-density outermost zones and ensure the stability of the star \citep{tyr17,tyr20a}.  This pressure, $P$, is parametrised by $P_{\mathrm{extra\_factor}}$,
\begin{equation}
    P = \frac{\tau g}{\kappa}\left(1 + P_{\textrm{extra}}\right),
\end{equation}
where
\begin{equation}
    P_{\textrm{extra}} = P_{\textrm{extra\_factor}}\frac{\kappa}{\tau} \ \frac{M}{L} \left(6\pi c G\right)^{-1}, 
\end{equation}
and $g = GM/R^2$, the surface gravity of the star, $\kappa$ and $\tau$ are the opacity and optical depth, and $M$ and $L$ are the mass interior to those layers and their luminosity, respectively.  We set $P_{\mathrm{extra\_factor}} = 1.36$ for our low accretion rate models, 0.01 - 0.001 \Ms\ yr$^{-1}$.

\section{SMS Evolution}

\begin{figure}
\centering
\includegraphics[width=0.45\textwidth]{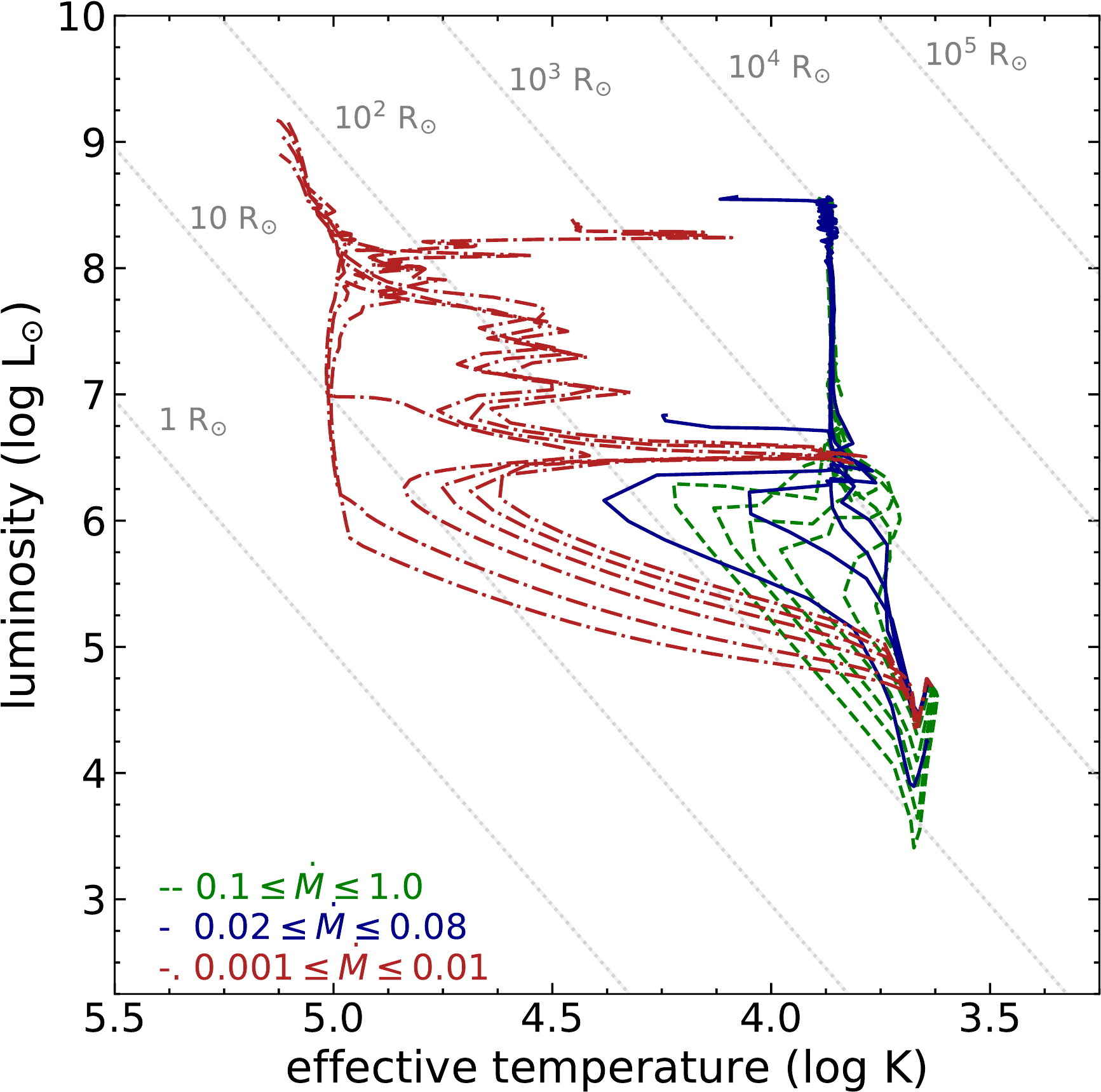}
\caption{Hertzsprung-Russell diagram of the evolution of all our SMSs, plotting every 15th output.  Accretion decades are separated by colour and line style: red dot-dashed lines are $10^{-3} - 10^{-2}$ \Ms\ yr$^{-1}$, blue solid lines are $10^{-2} - 10^{-1}$ \Ms\ yr$^{-1}$ and green dashed lines are $10^{-2} - 1.0$ \Ms\ yr$^{-1}$.}
\label{fig:HRmain}
\end{figure}

\begin{figure}
\centering
\includegraphics[width=84mm]{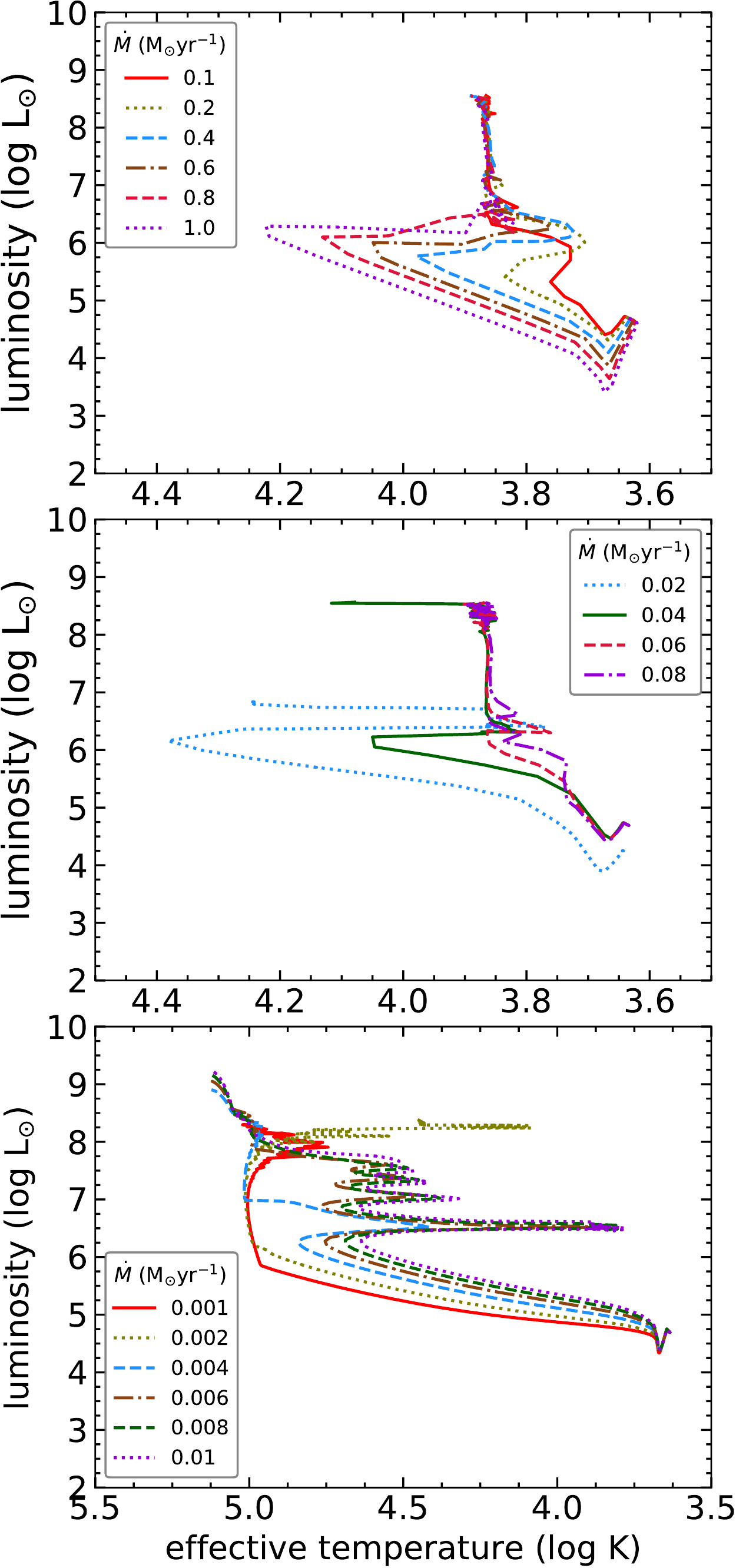}
\caption{Hertzsprung-Russell diagrams of the SMSs sectioned by accretion decade.}
\label{fig:HRsubplot}
\end{figure}

We show Hertzsprung-Russel (HR) plots of the evolution of our stars at accretion rates of 0.001 \Ms\ yr$^{-1}$ - 1 \Ms\ yr$^{-1}$ in Figures~\ref{fig:HRmain} and \ref{fig:HRsubplot} and Kippenhahn diagrams of their structures over time in Figures~\ref{fig:kipp_highAR} - \ref{fig:kipp_lowAR}.  How the stars evolve primarily depends on how accretion timescales, $t_{\textrm{acc}} = M/\dot{M}$, compare to Kelvin-Helmholz (KH) contraction times, $t_{\textrm{KH}} \approx GM^2/RL$, early in their lives where $\dot{M}$ and $L$ are the accretion rate and luminosity of the star, respectively. If the protostar gains enough mass during contraction it will have sufficient opacity in the outer layers to expand as a cool red SMS. If not, it will remain compact and hot and evolve along bluer tracks.  At first, our protostars are fully convective, with central temperatures below those required for deuterium burning.  They contract under accretion until their temperatures are high enough to ignite proton-proton (PP) burning.  

However, as shown by the early dips in all the HR tracks in Figure~\ref{fig:HRmain}, PP burning does not generate enough energy to halt contraction, which continues until the stars begin CNO burning, which is quickly followed by the triple alpha process.  The onset of the CNO cycle thus marks the beginning of stable nuclear burning in the star.  The PP chain raises central temperatures that drive changes in internal opacities that create a short-lived radiative core.  The mass at which the star forms a radiative core increases with accretion rate because the core acquires more mass during initial contraction.  However, CNO burning dramatically increases energy production so the core becomes convective to transport this energy outward more efficiently.  The stars eventually develop large convective cores and high-entropy outer envelopes like those in \citet{tyr17} and \citet{hle18b}. 

As in \citet{hle18b}, each star converges to a monotonic mass-luminosity evolution after an early period of reconfiguration that, as discussed above, follows either a cool red track or a hot blue one.  As shown in Figures~\ref{fig:HRmain} and \ref{fig:HRsubplot}, stars growing at rates above 0.01 \Ms\ yr$^{-1}$ - 0.02 \Ms\ yr$^{-1}$ branch off to evolve as red hypergiants while those below these rates become compact blue supergiants.  The transition from one regime to the other occurs at $\sim$ 0.02 \Ms\ yr$^{-1}$, as this SMS oscillates between red and blue phases.  As in \citet{hos13} and \citet{hle18b}, effective temperatures above this rate remain at $\sim$ 10$^4$ K even as they reach luminosities of $10^{9} - 10^{10}$ \Ls\ as the stars evolve along the Hayashi limit.  Below 0.01 \Ms\ yr$^{-1}$ the stars evolve onto the zero-age main sequence (ZAMS) and reach temperatures of up to $10^{5.5}$ K and luminosities of $10^8 - 10^{10}$ \Ls.  

The two tracks bifurcate during initial contraction, prior to any nuclear burning.  This transition can also be seen in the interiors of the stars shown in Figure~\ref{fig:kipp_midAR}.  Above 0.02 \Ms\ yr$^{-1}$ there are more radiation dominated structures in which less of the interior resides within the convective core.  At lower accretion rates the stars are almost fully convective, and most of the star is a burning core.  It is clear from Figure~\ref{fig:HRmain} that high accretion rate stars grow significantly in radius, and this is due to the fact that they gain mass more quickly than they can contract.  Low accretion rate stars contract more quickly than they gain mass so their radii decrease as they settle onto the ZAMS.  

\subsection{$\dot{M} =$ 0.1 - 1 \Ms\ yr$^{-1}$}

\begin{figure}
\centering
\includegraphics[width=79mm]{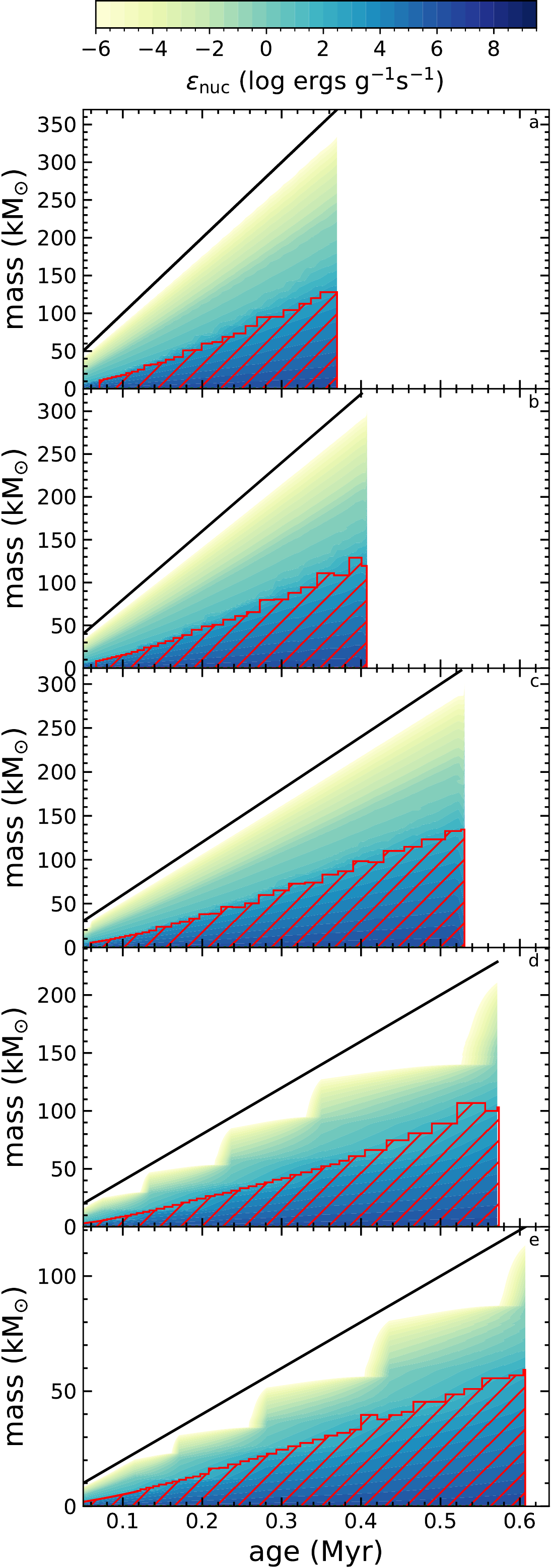}
\caption{Kippenhahn diagrams for SMSs with accretion rates of 1.0 \Ms\ yr$^{-1}$ - 0.2 \Ms\ yr$^{-1}$ (going from a to e), beginning at 10,000 \Ms\ during main burning. The colour map indicates nuclear burning rates, $\varepsilon$, in erg s$^{-1}$ gm$^{-1}$, increasing from blue to green to yellow. Black lines show the total mass at the star's current age. The red hatching marks convection regions, empty white-space regions have no mixing.}
\label{fig:kipp_highAR}
\end{figure}

As shown in the top panel of Figure~\ref{fig:HRsubplot}, these cool red stars evolve very quickly over a narrow range of effective temperatures.  As they grow in mass, they encounter opacity bumps that lead to excursions in the HR diagram until they converge to a direct track along the Hayashi limit.  Energy production by the PP, CNO and triple alpha chains causes the stars to grow in radius by a factor of ten during main sequence burning.  This expansion is driven by temperature sensitive H$^{-}$ opacity in their atmospheres, which leads to efficient transport of energy from the interior to the outer layers.  In Figure~\ref{fig:kipp_highAR}, the internal structures of these models change with decreasing accretion rate.  At lower accretion rates more of the stellar interior is subsumed by the convective core.  The staircase-like growth of the central convective zone is due to the finite resolution of our models and the limitations of our 1D prescription for convective instability.  

\subsection{$\dot{M} =$  0.02 - 0.1 \Ms\ yr$^{-1}$}

As shown in the middle panel of Figure~\ref{fig:HRsubplot}, most of the stars in this accretion range evolve in the same manner as our highest accretion rate stars, with cool red tracks and radial expansion after converging to the Hayashi limit.  The transition from red to blue SMSs occurs at the lower end of this range, as we discuss in greater detail in Section 3.6.  The star accreting at 0.02 \Ms\ yr$^{-1}$ alternates between red and blue tracks because densities and temperatures in its atmosphere at times are like those in high accretion rate models, but the star does not gain sufficient mass over time to sustain them and periodically contracts back down to a bluer state.  These transitions also cause \texttt{MESA} to readjust its timesteps upon excursion from one effective temperature to another.  

\begin{figure}
\centering
\includegraphics[width=80mm]{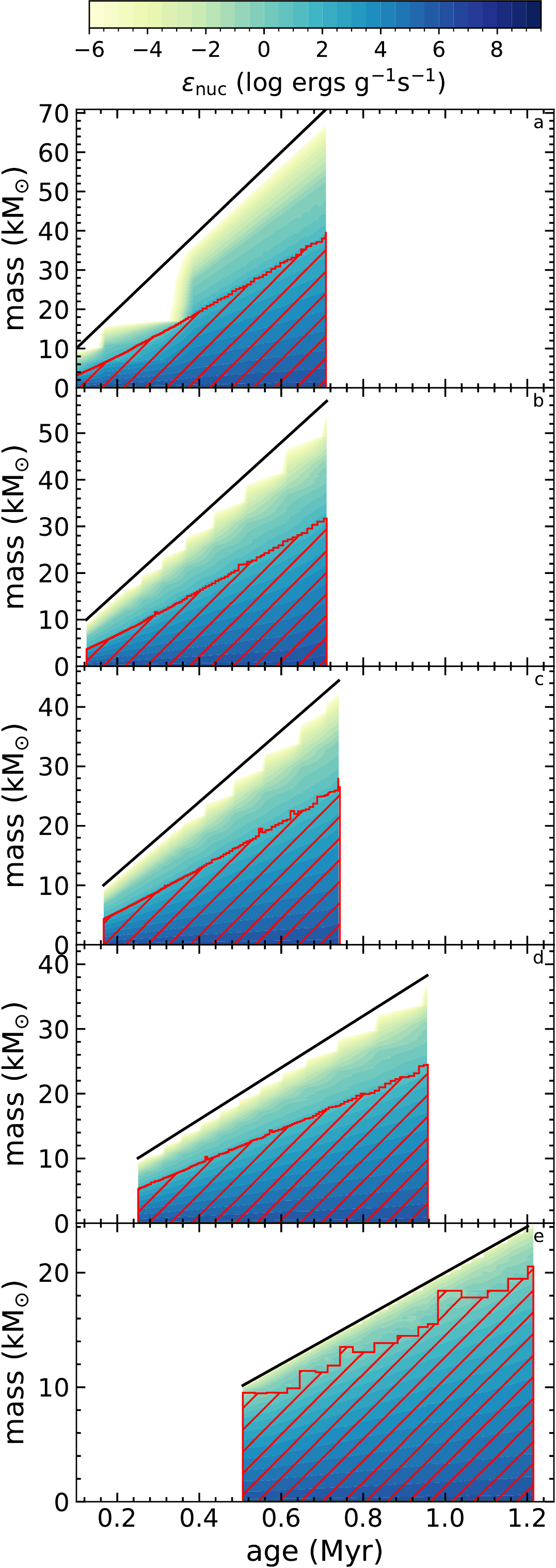}
\caption{Kippenhahn diagrams for SMSs with accretion rates of 0.1 \Ms\ yr$^{-1}$ - 0.02 \Ms\ yr$^{-1}$ going from a to e.}
\label{fig:kipp_midAR}
\end{figure}

The large swings in the structure of the 0.02 \Ms\ yr$^{-1}$ star prematurely end its run but adjustments to the pressure at its surface allow it to be evolved to the end of main sequence burning, as discussed in Section 3.6.  There we show that the star continues to cycle between red and blue states before finally settling onto a blue track and becoming hot.  As shown in Figure~\ref{fig:kipp_midAR}, a significant fraction of the mass of the 0.02 \Ms\ yr$^{-1}$ star resides within its convective core.  At lower accretion rates the percentage of the star that is convective rises.  Nuclear energy generation in the convective core is mostly due to CNO, triple alpha and nitrogen burning.  Outside the convective zone, weak energy generation proceeds by the PP chain but is dwarfed by core burning. It is clear from Figure~\ref{fig:kipp_midAR} that the 0.02 \Ms\ yr$^{-1}$ star is evolving towards the low accretion rate regime, as this model is comparatively much older once it reaches 10$^4$ \Ms.

\subsection{$\dot{M} =$ 0.001 - 0.01 \Ms\ yr$^{-1}$}

\begin{figure}
\centering
\includegraphics[width=80mm]{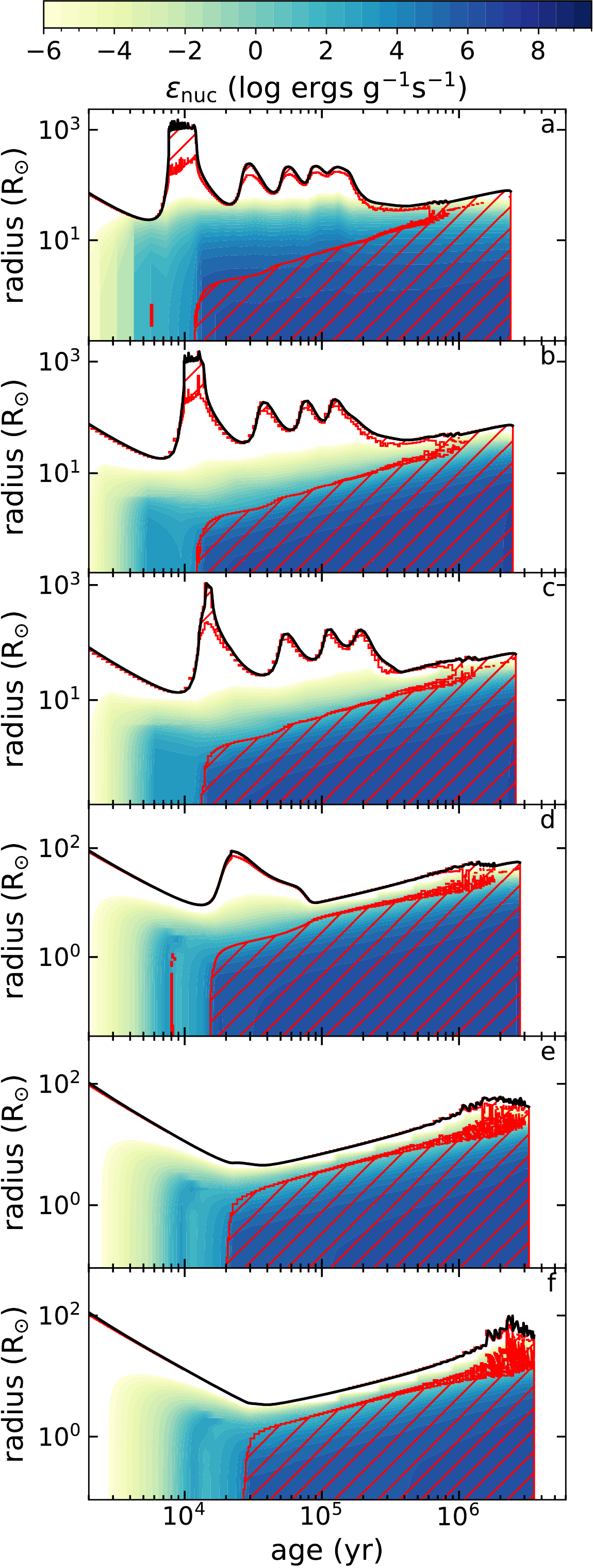}
\caption{Kippenhahn diagrams for SMS with accretion rates of 0.01 \Ms\ yr$^{-1}$ - 0.001 \Ms\ yr$^{-1}$ (labelled a to f) showing radius with age, from a few thousand years into the pre-mainsequence to the onset of post-main sequence burning.}
\label{fig:kipp_lowAR}
\end{figure}

HR tracks for the SMSs in the lowest accretion range are shown in the bottom panel of Figure~\ref{fig:HRsubplot}.  They evolve as hot blue stars that can be more than a hundred times smaller than red SMSs of comparable mass.  At early stages in their evolution these stars have similar radii as they approach the ZAMS.  Just before reaching the post-main sequence, they begin to expand as they deplete their central hydrogen supply, with those with the highest rates having the largest radii and luminosities.  We show the internal structures of these stars in Figure~\ref{fig:kipp_lowAR}.  Most of the mass of these compact blue stars resides in their convective cores.  As at high accretion rates, the staircase-like growth of the central convective zone is again due to the finite resolution of our models and the limitations of our 1D prescription for convective instability.  

\begin{figure}
\centering
\includegraphics[width=0.47\textwidth]{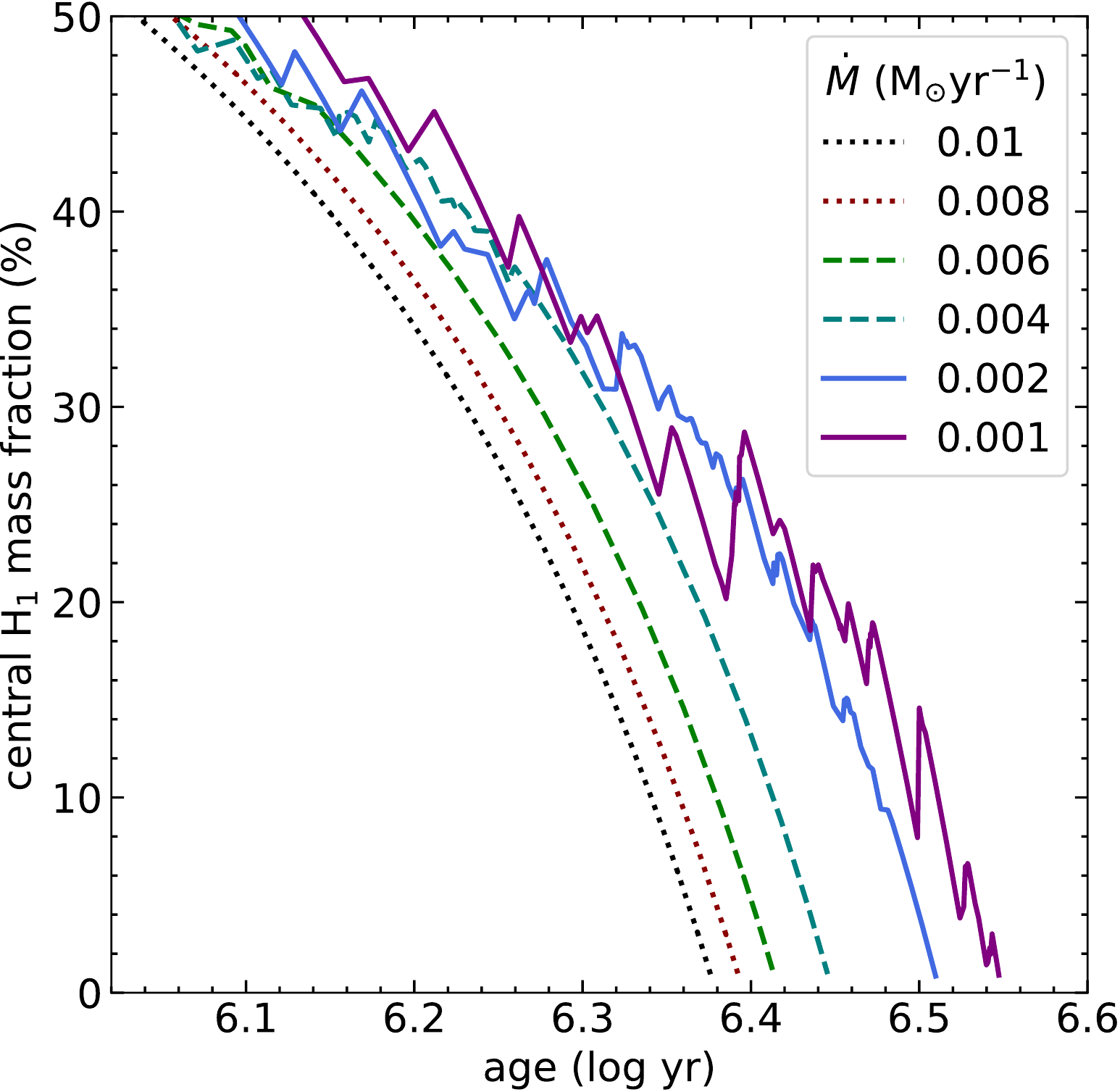}
\caption{Central hydrogen mass fractions in the stars in the lowest accretion decade.}
\label{fig:inges}
\end{figure}

In principle, convection in these stars could periodically replenish their cores with H from higher mass coordinates.  If such an event were to occur after core H mass fractions fall below 1\% our models would underestimate the true lifetime and final mass of the star because the ingestion of fresh H into the core would have allowed the star to continue to burn after we terminated the run.  However, as we show in Figure~\ref{fig:inges}, such ingestion events only occur near the end of the life of the star at 0.001 \Ms\ yr$^{-1}$.  It is not possible at present to evolve this star further because of numerical instabilities that arise in the run during the contraction of the core and the expansion of the star as it becomes fully convective at this stage, so the final mass of this star should be taken to be a lower limit.

\subsection{Collapse}

\begin{figure} 
\centering
\begin{tabular}{c}
\includegraphics[width=0.48\textwidth]{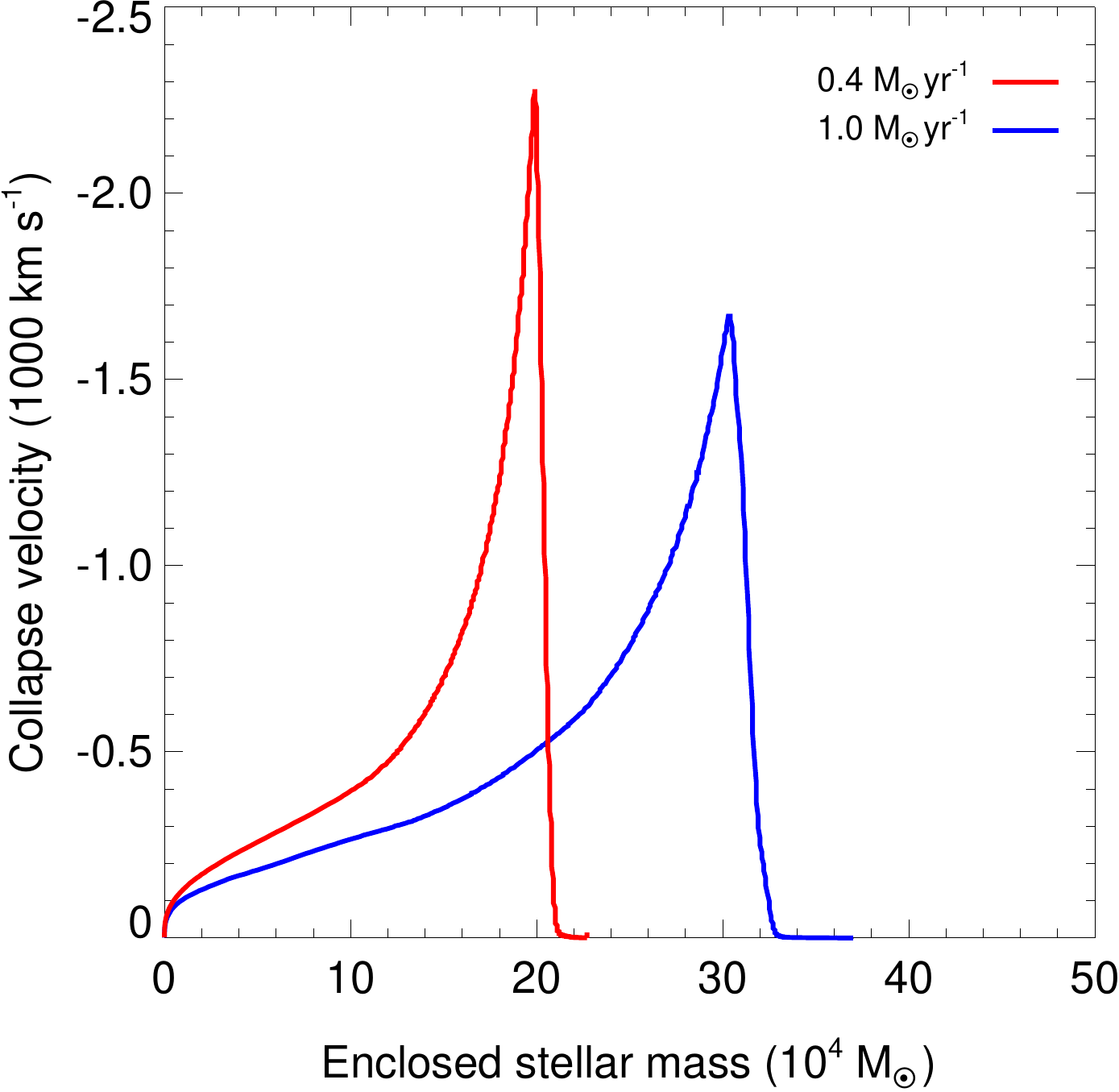}  \\
\includegraphics[width=0.48\textwidth]{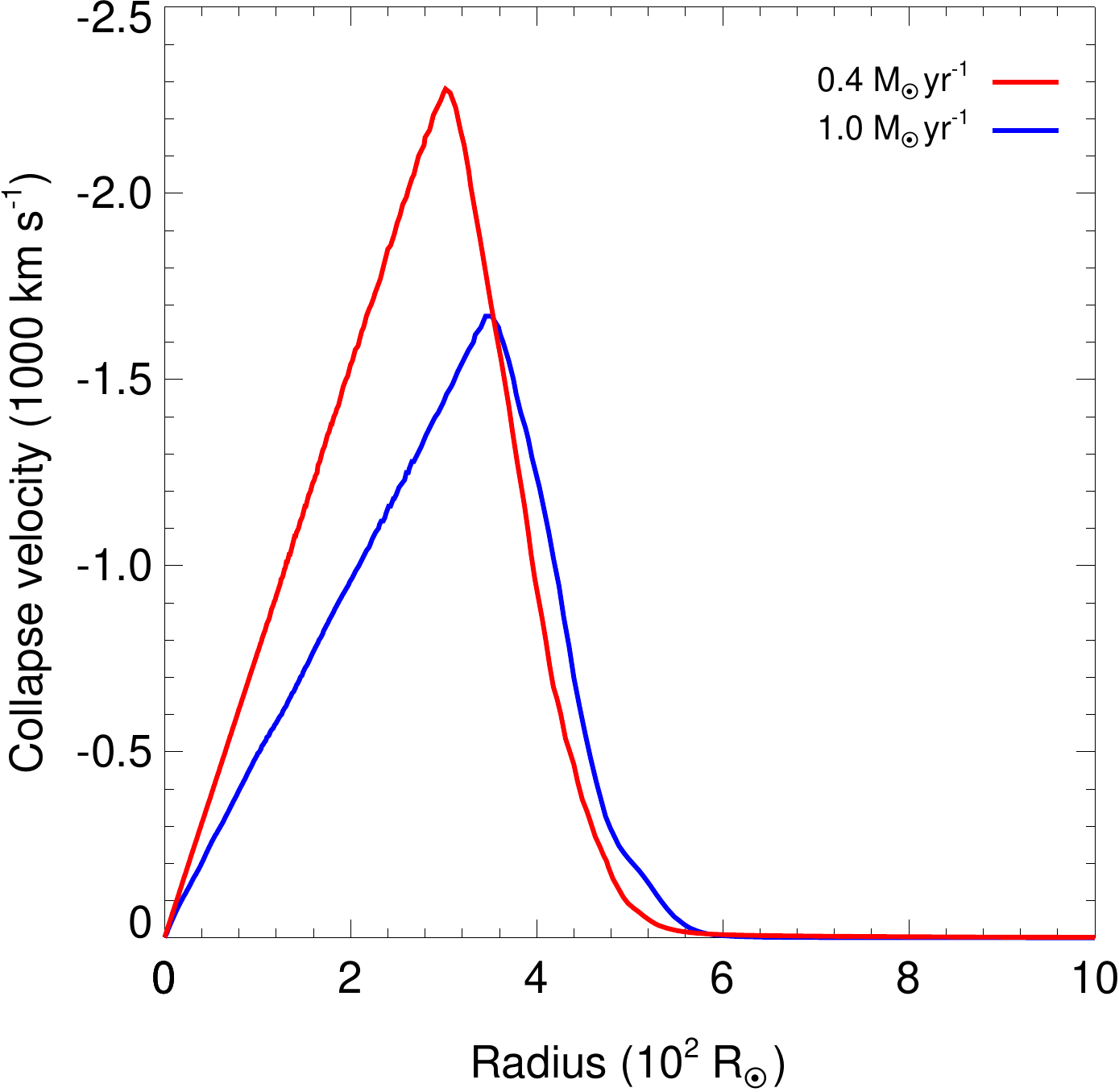}  \\
\end{tabular}
\caption{Internal velocity profiles of the 0.4 \Ms\ yr$^{-1}$ and 1.0 \Ms\ yr$^{-1}$ stars at collapse vs mass coordinate (top) and radius (bottom).
\label{fig:coll}}
\end{figure}

As discussed in greater detail in the next section, SMSs growing at 0.02 - 1 \Ms\ yr$^{-1}$ become unstable during hydrogen burning because of changes in central temperatures and densities induced by pulsations due to the post-Newtonian instability and collapse before the end of the main sequence.  Once collapse proceeds, infall rates in the core rapidly reach 1000 km s$^{-1}$ as shown in Figure~\ref{fig:coll}.  Stars growing at 0.02 - 0.1 \Ms\ yr$^{-1}$ evolve further along the main sequence before collapse, with the 0.02 \Ms\ yr$^{-1}$ SMS reaching final central hydrogen fractions of 0.12. The 0.01 \Ms\ yr$^{-1}$ SMS marks the transition to low accretion rate evolution in which the star reaches the central hydrogen fraction cutoff of 0.01 and develops a helium core as it advances to later stages of burning.  Stars that grow at 0.001 - 0.01 \Ms\ yr$^{-1}$ reach the post-main sequence, evolving in a similar manner as massive Pop III stars.  They never encounter the post-Newtonian instability and, although we do not model it here, are expected to collapse during He, C or O burning.  Accretion rates of 0.01 \Ms\ yr$^{-1}$ - 0.02 \Ms\ yr$^{-1}$ thus mark a key divide in SMS evolution: whether they evolve as cool red hypergiants or compact blue supergiants and if they collapse via the GRI or because of core fuel depletion.

\subsection{Hot CNO cycle}

\begin{figure}
\centering
\includegraphics[width=0.48\textwidth]{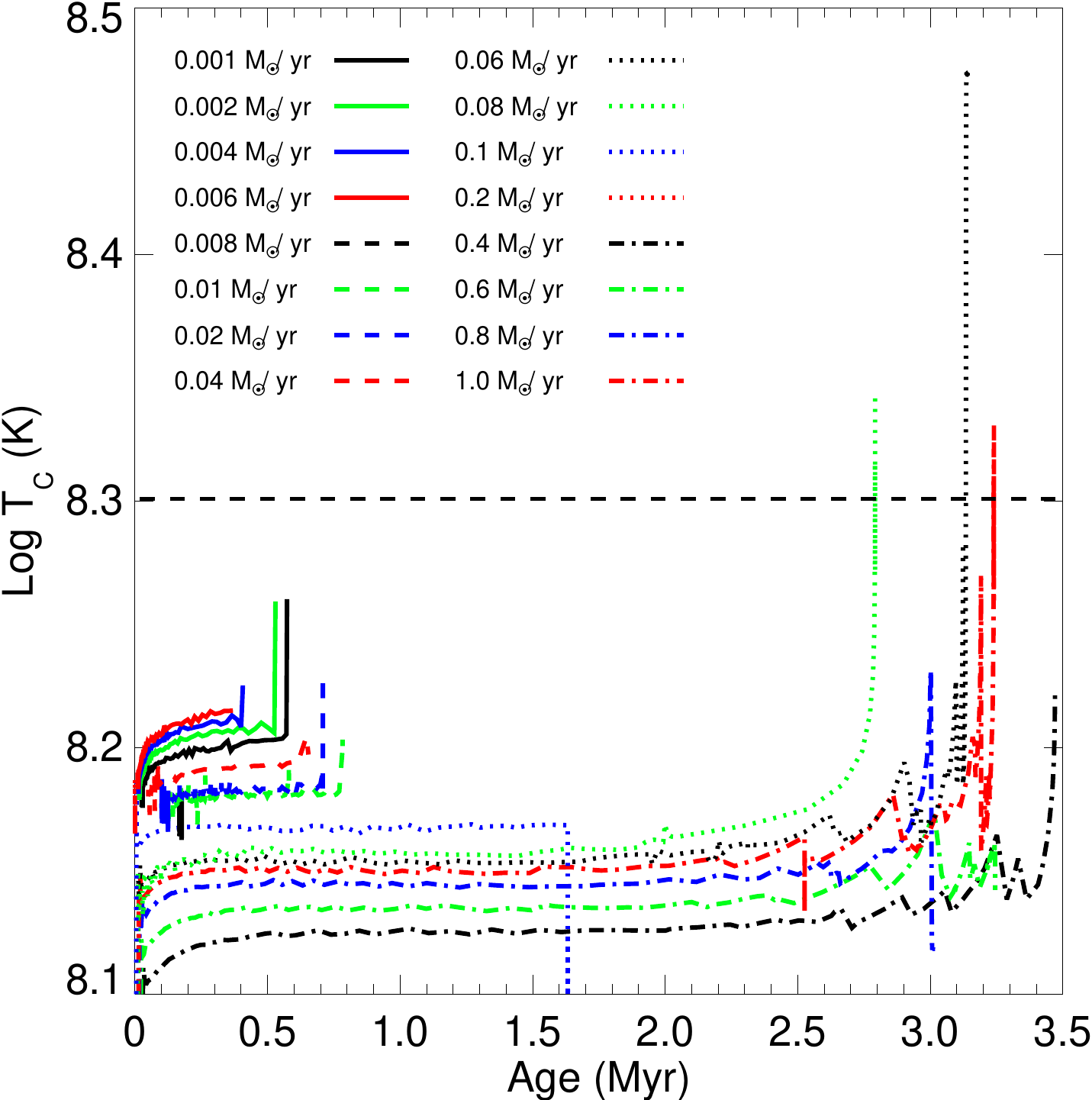}
\caption{Central temperatures for all the stars over their evolution.  The dashed horizontal line marks $T =$ $2 \times 10^8$ K, at which energy production due to rapid proton captures in the hCNO cycles begin to be important.}
\label{fig:Tc}
\end{figure}

\begin{figure*} 
\begin{center}
\begin{tabular}{cc}
\includegraphics[width=0.48\textwidth]{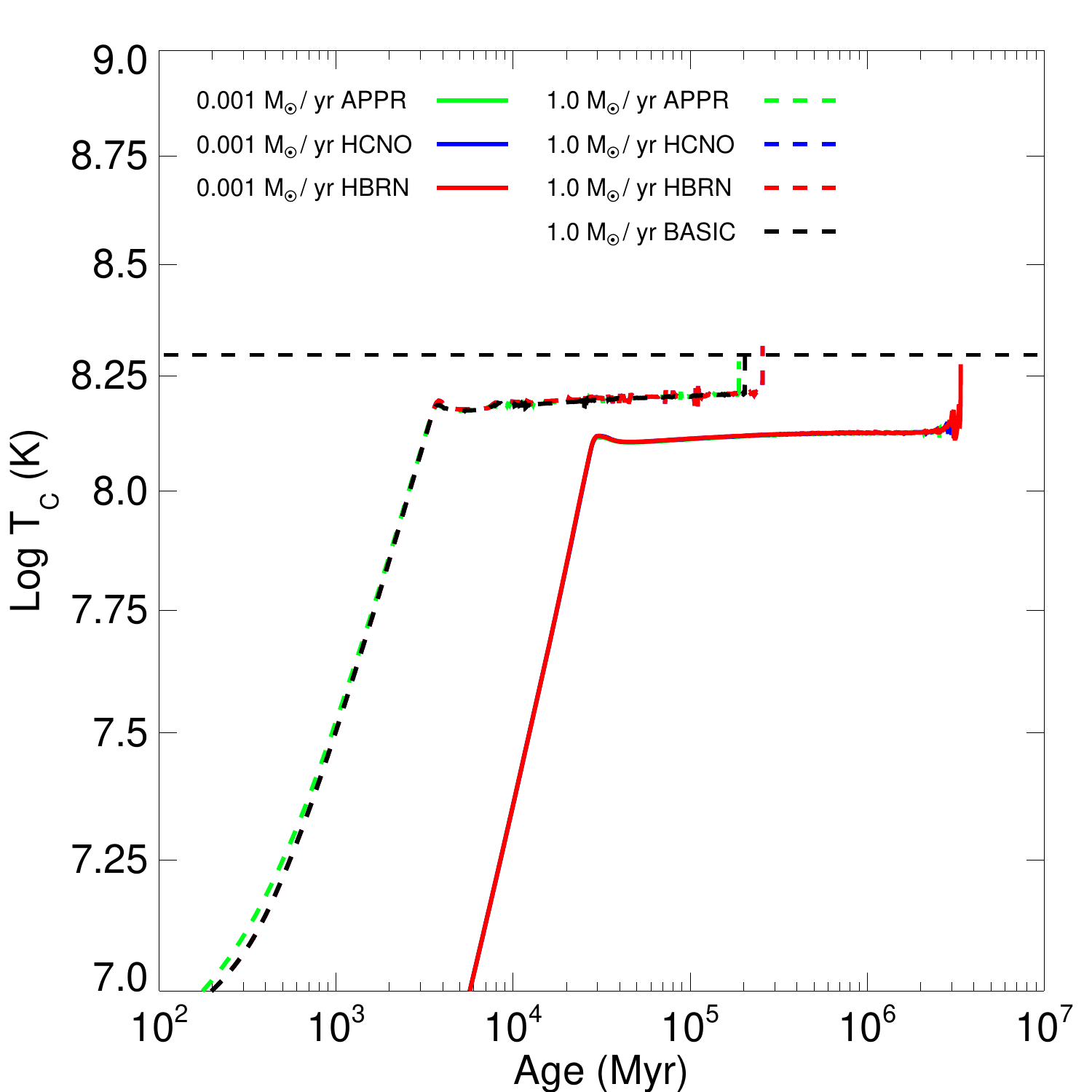}  &
\includegraphics[width=0.48\textwidth]{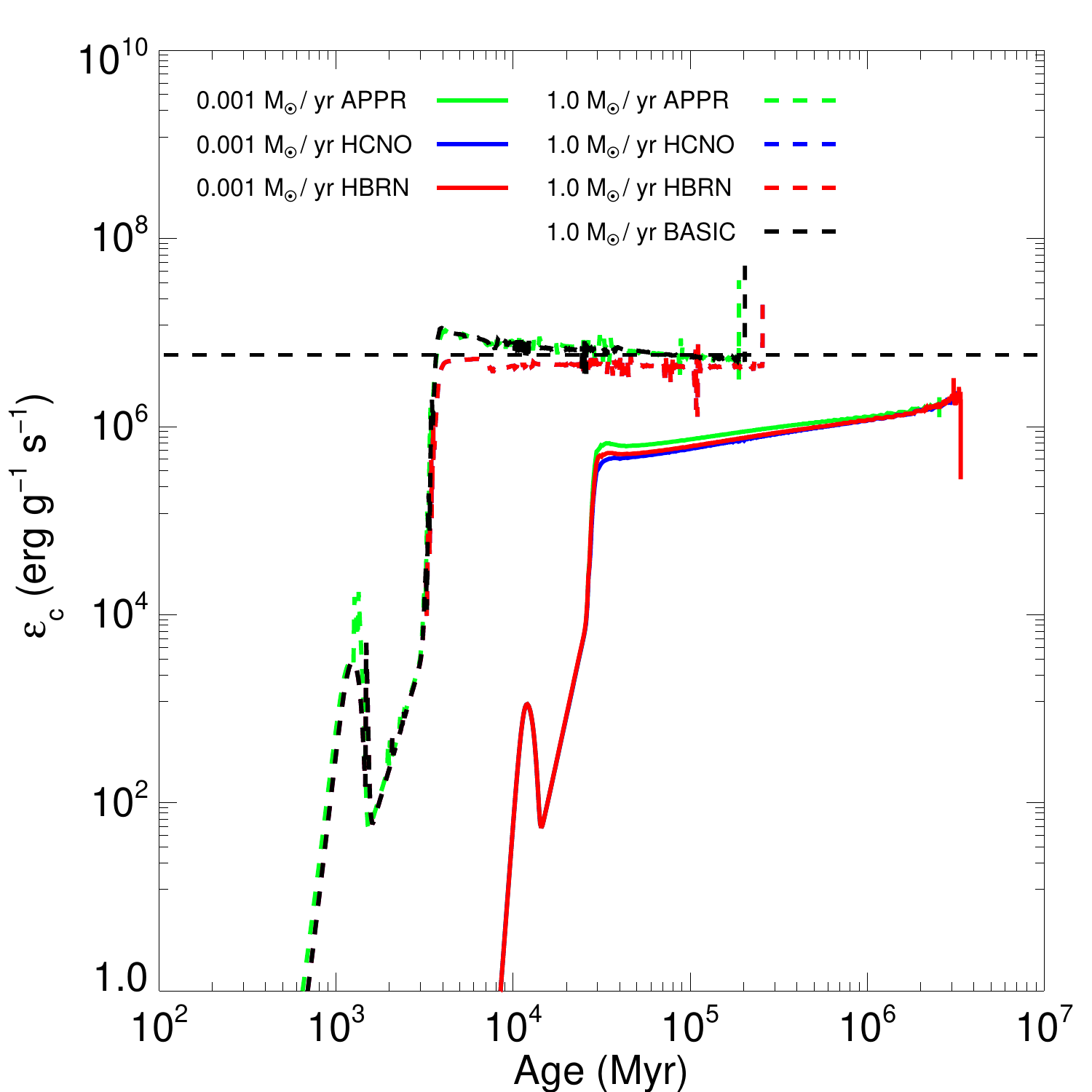}  
\end{tabular}
\end{center}
\caption{Comparison of central temperatures $T_c$ (left) and nuclear energy production rates  $\epsilon_c$ (right) for the 0.001 \Ms\ yr$^{-1}$ and 1.0 \Ms\ yr$^{-1}$ stars with the APPROX21, HCNO, 44-isotope HBURN and 8-isotope BASIC nuclear reaction networks.}
\vspace{0.1in}
\label{fig:TCNO} 
\end{figure*}

If core temperatures in SMSs reach $2 \times 10^8$ K while still at high central hydrogen fractions, energy generation due rapid proton (rp) captures can rival that of the CNO cycle and become several hundred times greater at temperatures above $5 \times 10^8$ K because CNO reaction rates are limited by the half-life of one of its beta decays \citep[$\beta$-limited CNO;][]{fuller86}.  However, as shown in Figure~\ref{fig:Tc} core temperatures calculated with the 21-isotope APPROX network never exceed $2 \times 10^8$ K in our stars except for brief episodes late in their lives when central hydrogen fractions are low.  In principle, the inclusion of more isotopes and rp captures (the hot CNO, or hCNO, cycle) could have produced larger core temperatures than those in our models.  We ran the 0.001 \Ms\ yr$^{-1}$ and 1.0 \Ms\ yr$^{-1}$ stars with the 44-isotope HBURN network with one hCNO cycle and the reduced HCNO network with just the p-p chain, triple alpha chain, CNO and hCNO cycles to compare core temperatures and energy production rates, which are plotted in Figure~\ref{fig:TCNO} (we also ran the 1.0 \Ms\ yr$^{-1}$ star with the 8-isotope BASIC network, which has the p-p chain, triple alpha chain, and CNO cycle, for comparison).  Unlike non-accreting SMS models, which exhibit core temperatures at which the hCNO cycle becomes important \citep[e.g.,][]{fuller86,nag22}, we find that the inclusion of the hCNO cycle in our models does not produce central temperatures at which it becomes important.  Indeed, there is little difference in core temperatures with the three networks.  The inclusion of more isotopes stabilizes energy production in the core at earlier times but results in the same evolution and final mass for the star.

\subsection{0.02 \Ms\ / yr SMS Atmosphere Effects}

\begin{figure}
\centering
\includegraphics[width=85mm]{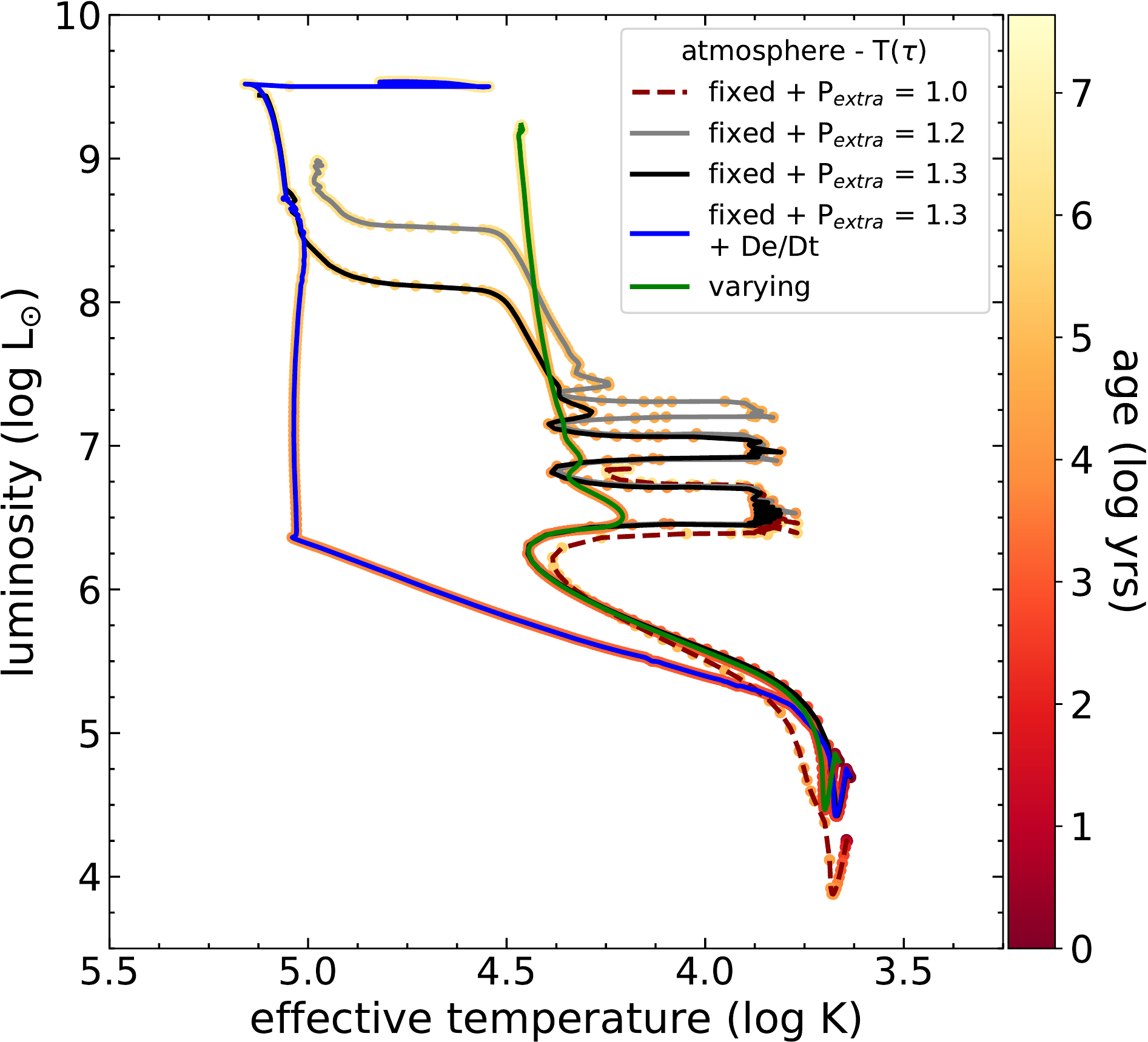}
\caption{HR diagram for the 0.02 \Ms\ yr$^{-1}$ SMS. 'Varying' is a $T(\tau)$ atmosphere with a variable opacity and 'fixed' is a $T(\tau)$ atmosphere with a uniform opacity. $P_{\mathrm{extra}}$ is related to the pressure imposed on the surface of the star as discussed in Section~\ref{section:method} and De/Dt is an alternative formulation of the energy equation with improved numerical energy conservation \citep{Paxton19}. Our original 0.02 \Ms\ yr$^{-1}$ SMS model is shown in the dashed red line.}
\label{fig:002atmosphere}
\end{figure}

As shown in Figure~\ref{fig:HRsubplot}, the 0.02 \Ms\ yr$^{-1}$ star alternates between red and blue tracks, suggesting that this is the accretion rate at which SMSs transition from blue to red.  Its evolution is sensitive to the choice of atmosphere and surface pressure. We evolved the star with two versions of the standard \texttt{MESA} T($\tau$) atmosphere, one in which the opacity is constant throughout the atmosphere ('fixed', which was used for all the models in our study) and one in which it varies in a manner that is consistent with the local pressure and temperature ('varying'). For the fixed case we tested two extra pressures at the surface of the star that are set by $P_{\textrm{extra\_factor}} =$ 1.2 and 1.3, as discussed at the end of Section 2.  We also consider an alternate formulation for the energy equation, De/Dt, that better conserves energy \citep{Paxton19}. Figure \ref{fig:002atmosphere} shows that changing only the pressure boundary for this star can lead to either a hot blue or cool red SMS after the onset of main sequence burning. The varying atmosphere option stabilises the thermal oscillations and leads to a track that is intermediate to the cool red and hot blue regimes. Testing these options in our other models produced much smaller deviations in evolution in the 0.01 \Ms\ yr$^{-1}$ and 0.03 \Ms\ yr$^{-1}$ stars and few if any in the others, confirming that 0.02 \Ms\ yr$^{-1}$ is the transitional accretion rate for rapidly accreting SMSs.

During the protostellar phase the 0.02 \Ms\ yr$^{-1}$ SMS initially contracts on a shorter timescale than it can accrete and evolves as a hot blue star. After central nuclear burning begins, the choice of atmosphere leads to deviations in evolution by sending the outer layers of the star into regions of $T$ - $\rho$ space that favor or suppress H$^{-}$ formation, whose opacity intercepts energy from the core of the star and can cause the outer layers to expand along the Hayashi limit. In the original model, the surface layers of the star were at temperatures and densities that were marginally favorable to H$^{-}$ formation.  As the star expanded its outer layers migrated into regions of $T$ - $\rho$ space in which H$^{-}$ tended to be destroyed (likely because falling densities decreased the two-body H$^{-}$ formation rate) and the star again began to contract to a hotter, bluer phase.  

Increasing the pressure on the outer boundary moves these layers into regions in $T$ - $\rho$ space that are less favorable to H$^{-}$ formation and cause the star to settle onto a blue track at earlier times.  Excursions between red and blue states are dampened in the varying case and the star settles onto an intermediate track at early times because the temperature structure of the atmosphere can respond quickly to changes in opacity due to  those in density.  The run with De/Dt settles onto the hot blue track at early times because the energy equation produces consistently higher surface temperatures that prevent H$^{-}$ from forming and expanding the star.  The varying case suggests that the oscillations of this star in the HR diagram are probably not in reality as large as the other test runs suggest but, as noted earlier, this and the De/Dt option had little effect of the evolution of the other stars.

\section{Discussion}

High accretion rate stars encounter pulsations due to the GRI at masses above a few tens of thousands of solar masses.  These pulsations are initially mild, and can drive shocks into the core with speeds of a few tens of km s$^{-1}$. The shocks only induce small changes in central densities and temperatures from which the star can easily recover, unless code time steps that are shorter than nuclear burning times but too long for the hydrodynamics to return accurate velocities lead to unphysically large infall speeds, as discussed earlier.  However, as the star grows in mass the pulsations become more violent and drive shocks into the core with velocities of hundreds of km s$^{-1}$.  These shocks elevate central temperatures and densities that in turn exacerbate the GRI and lead to even stronger pulsations.  The stars reach their final masses when a pulsation finally triggers collapse.  

In tests with no hydrodynamics, there are no shocks to induce changes in central densities or temperatures so   the model ends when the GRI causes enough softening of the EOS in the core, usually at significantly larger masses as discussed below.  We note that both high and low accretion rate stars could be prone to other types of pulsations that did not appear in our models because they had characteristic timescales that were shorter than the time steps taken by our simulations.  Further study is required to determine if they occur and what impact they have on the evolution of the star, such as if they can lead to collapse at lower masses than those due to the GRI.

\label{section:finalmass}

\begin{figure}
\centering
\includegraphics[width=0.45\textwidth]{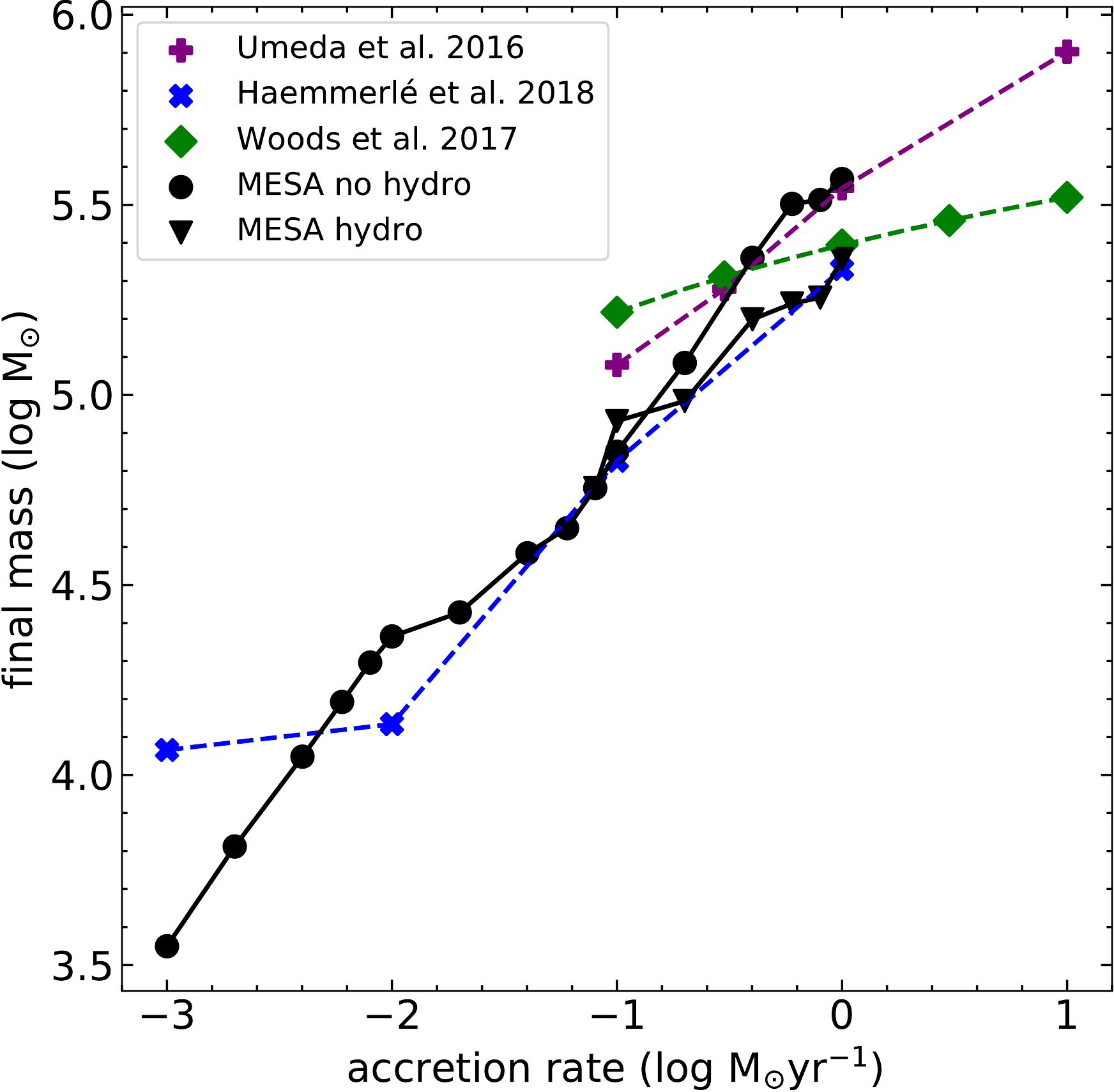}
\caption{Final masses of SMSs as a function of accretion rate compared to those in \citet{tyr17}, \citet{hle18b}, and \citet{um16}. The  triangles and circles in black are \texttt{MESA} models with and without implicit hydrodynamics, respectively.}
\label{fig:finalmass}
\end{figure}
\label{section:hydro}

We show final masses for the stars in Figure~\ref{fig:finalmass} with those from previous studies at the same accretion rates.  As in earlier work, our final masses rise monotonically with accretion rate, and above 0.01 \Ms\ yr$^{-1}$ they fall in between those of \citet{hle18b} and \citet{tyr17}.  We modeled stars growing at above 0.01 \Ms\ yr$^{-1}$ with and without hydrodynamics to test its effects on their evolution and final masses.  From 0.01 \Ms\ yr$^{-1}$ - 0.1 Ms\ yr$^{-1}$ they converge to essentially the same mass.  Stars growing at $\ge$ 0.1 \Ms\ yr$^{-1}$ without hydrodynamics have consistently larger final masses that are closer to those of \citet{um16} because they do not manifest the pulsations that would have collapsed the star at earlier times and because the code can take larger time steps that can mitigate other effects of the GRI.  

As noted earlier, at accretion rates of 0.01 \Ms\ yr$^{-1}$ stars begin to collapse because of hydrogen depletion while at rates of 0.02 \Ms\ yr$^{-1}$ they collapse via the GRI.  What are the relative roles of the two processes in collapse over this range?  Either one can lead to excessive changes in central densities and temperatures that drive catastrophic reductions in time step that signify the death of the star.  When we ran the 0.02 \Ms\ yr$^{-1}$ star without post-Newtonian corrections it evolved further into hydrogen depletion and encountered numerical difficulties as it entered the post main sequence, but it did not develop large infall velocitiies we would associate with collapse. Likewise, when we ran the 0.01 \Ms\ yr$^{-1}$ star without post-Newtonian corrections, the star struggled through hydrogen depletion and then collapsed.  These two tests indicate that the GRI was primarily to blame for the collapse of the first star and depletion of core hydrogen caused the collapse of the second.  Both likely contribute to collapse at 0.01 \Ms\ yr$^{-1}$ - 0.02 \Ms\ yr$^{-1}$.

Stars accreting at the lowest decade in rate have high temperatures and large ionizing UV fluxes that could in principle disperse the flows that create them, raising the question of whether such stars could grow at these rates in the first place.  \citet{latif21a} studied the growth of such stars and included the effects of their radiation on their inflows.  They found that they could continue to grow at rates in the lowest decade we considered.  In reality, no SMS grows at uniform accretion rates, and even cool, red stars whose radiation has no effect on accretion are subject to time-dependent cosmological flows.  However, studies of SMSs have begun to consider their evolution in such flows and find that they evolve in much the same way as at constant accretion rates \citep{tyr21a,tyr22a}.

As mentioned in the Introduction, SMS are not expected to lose mass to line-driven winds because of the low opacity of primordial gas.  However, the stars soon become highly convective and heavier elements produced by the core could be dredged up and drive winds later in the life of the star.  But it is unlikely that these winds would lead to significant mass loss because they would be modest and probably be overcome by the ram pressure of the infalling gas.  Consequently, the final mass of the star will just be the mass it accrued over its lifetime.

SMSs in a narrow range of masses around 55,000 \Ms\ in which accretion has subsided and the star has thermally relaxed \citep[e.g.,][]{tyr20a} have been found to explode in previous studies \citep{montero12,wet12a,jet13a,wet13b,wet13d,chen14b,nag21,mor21,nag22}.  However, there could be other observational signatures of SMS collapse even if they do not explode.  For example, if the star rotates and core collapse drives a jet that pierces its outer layers, it could imprint distinctive nucleosynthetic patterns on surrounding gas that could later be found in the atmospheres of low mass stars forming in it \citep[e.g.,][]{jet09b}.  Collapse could also produce a strong neutrino signal that is far brighter than that unleashed by conventional core-collapse SNe, but it would still only be detectable at Mpc distances \citep{sfh98,mun21,nag21}.

Although we do not follow the collapse of our stars to late times, \citet{nag21} found that event horizon formation during the collapse of a 10$^4$ \Ms\ SMS intially encloses 40 - 50 \Ms\ of the core of the star, evoking the possibility of the birth of a quasi-star \citep[e.g.,][]{begel08,vb10}.  In this picture, X-rays from the BH support the rest of the star from prompt collapse and form a stable envelope that would appear to be a cool, red giant star to an external observer.  Such stars can grow to $\sim$ 10$^6$ \Ms\ before the BH becomes so massive that a hydrostatic envelope is no longer possible.  However, even if the BH initially had similar masses in our stars it is unlikely that X-rays could halt the collapse of the star because so much of its mass is falling inward at velocities above several hundred km s$^{-1}$, as shown in Figure~\ref{fig:coll}.  Consequently, DCBHs are born with the mass at which their progenitors die.

\section{Conclusion}

We have carried out a systematic study of the evolution of rapidly accreting SMS with the publicly-available stellar evolution code \texttt{MESA}. Our grid of models spans 3 decades in accretion rate that bracket the range expected for primordial environments conducive to SMS formation.  We find that SMSs evolve along two different pathways, as cool red hypergiants or compact blue supergiants, at accretion rates above and below 0.01 $\leq \dot{M} \leq$ 0.02 \Ms\ yr$^{-1}$, respectively.  This range also marks the transition between stars collapsing because of the depletion of core fuel after the end of the main sequence at low rates and collapse via the GRI during the main sequence at higher rates.

We also find that hydrodynamics is crucial to capturing how the GRI causes the death of the star at higher accretion rates: triggering pulsations that eventually lead to its collapse.  Without hydrodynamics, the GRI still leads to the collapse of the star but at significantly higher masses by softening the EOS in the core and triggering ingestion events that raise central densities and temperatures and destabilise the core.  High accretion rate models with hydrodynamics encounter fatal unstabilities at lower final masses before ingestion events occur.  When our SMS models reach collapse, large infall velocities enclose most of the mass of the star so it will go into the BH soon after birth, preventing the formation of a quasistar that could create DCBHs of up to 10$^6$ \Ms\ \citep[e.g.,][]{begel08}. Our results are broadly consistent with previous, more sparsely-sampled accretion rates \citep{tyr17,hle18b} and, critically, provide a framework for future SMS modelling efforts with open-source tools.

\section*{Acknowledgements}

We thank Bill Paxton and the \texttt{MESA} community for valuable discussions that made this work possible. N.P.H. acknowledges funding from the European Research Council for the Horizon 2020 ERC Consolidator Grant project ICYBOB, grant number 818940. D.J.W. was supported by the Ida Pfeiffer Professorship at the Institute of Astrophysics at the University of Vienna and by STFC New Applicant Grant ST/P000509/1. T.E.W. acknowledges support from the NRC-Canada Plaskett Fellowship. 
\textit{Software:} \texttt{MESA} \citep{Paxton15,Paxton19}, MESASDK 20.12.1 \citep{mesasdk} and py\_mesa\_reader \citep{pymesa}.  

\section{Data Availability Statement}

The data in this study will be made available upon request to the corresponding author.

\bibliographystyle{mnras}
\bibliography{refs} 

\begin{thebibliography}{}
\makeatletter
\relax
\def\mn@urlcharsother{\let\do\@makeother \do\$\do\&\do\#\do\^\do\_\do\%\do\~}
\def\mn@doi{\begingroup\mn@urlcharsother \@ifnextchar [ {\mn@doi@}
  {\mn@doi@[]}}
\def\mn@doi@[#1]#2{\def\@tempa{#1}\ifx\@tempa\@empty \href
  {http://dx.doi.org/#2} {doi:#2}\else \href {http://dx.doi.org/#2} {#1}\fi
  \endgroup}
\def\mn@eprint#1#2{\mn@eprint@#1:#2::\@nil}
\def\mn@eprint@arXiv#1{\href {http://arxiv.org/abs/#1} {{\tt arXiv:#1}}}
\def\mn@eprint@dblp#1{\href {http://dblp.uni-trier.de/rec/bibtex/#1.xml}
  {dblp:#1}}
\def\mn@eprint@#1:#2:#3:#4\@nil{\def\@tempa {#1}\def\@tempb {#2}\def\@tempc
  {#3}\ifx \@tempc \@empty \let \@tempc \@tempb \let \@tempb \@tempa \fi \ifx
  \@tempb \@empty \def\@tempb {arXiv}\fi \@ifundefined
  {mn@eprint@\@tempb}{\@tempb:\@tempc}{\expandafter \expandafter \csname
  mn@eprint@\@tempb\endcsname \expandafter{\@tempc}}}

\bibitem[\protect\citeauthoryear{{Agarwal}, {Smith}, {Glover}, {Natarajan}  \&
  {Khochfar}}{{Agarwal} et~al.}{2016}]{agarw15}
{Agarwal} B.,  {Smith} B.,  {Glover} S.,  {Natarajan} P.,   {Khochfar} S.,
  2016, \mn@doi [\mnras] {10.1093/mnras/stw929}, \href
  {http://adsabs.harvard.edu/abs/2016MNRAS.459.4209A} {459, 4209}

\bibitem[\protect\citeauthoryear{{Aird} et~al.,}{{Aird} et~al.}{2013}]{athena}
{Aird} J.,  et~al., 2013, arXiv e-prints, \href
  {https://ui.adsabs.harvard.edu/abs/2013arXiv1306.2325A} {p. arXiv:1306.2325}

\bibitem[\protect\citeauthoryear{{Alvarez}, {Wise}  \& {Abel}}{{Alvarez}
  et~al.}{2009}]{awa09}
{Alvarez} M.~A.,  {Wise} J.~H.,   {Abel} T.,  2009, \mn@doi [\apjl]
  {10.1088/0004-637X/701/2/L133}, \href
  {http://adsabs.harvard.edu/abs/2009ApJ...701L.133A} {701, L133}

\bibitem[\protect\citeauthoryear{{Appenzeller} \& {Fricke}}{{Appenzeller} \&
  {Fricke}}{1972}]{af72a}
{Appenzeller} I.,  {Fricke} K.,  1972, \aap, \href
  {http://adsabs.harvard.edu/abs/1972A%26A....18...10A} {18, 10}

\bibitem[\protect\citeauthoryear{{Ardaneh}, {Luo}, {Shlosman}, {Nagamine},
  {Wise}  \& {Begelman}}{{Ardaneh} et~al.}{2018}]{ard18}
{Ardaneh} K.,  {Luo} Y.,  {Shlosman} I.,  {Nagamine} K.,  {Wise} J.~H.,
  {Begelman} M.~C.,  2018, \mn@doi [\mnras] {10.1093/mnras/sty1657}, \href
  {http://adsabs.harvard.edu/abs/2018MNRAS.479.2277A} {479, 2277}

\bibitem[\protect\citeauthoryear{{Ba{\~n}ados} et~al.,}{{Ba{\~n}ados}
  et~al.}{2018}]{ban18}
{Ba{\~n}ados} E.,  et~al., 2018, \mn@doi [\nat] {10.1038/nature25180}, \href
  {http://adsabs.harvard.edu/abs/2018Natur.553..473B} {553, 473}

\bibitem[\protect\citeauthoryear{{Baraffe}, {Heger}  \& {Woosley}}{{Baraffe}
  et~al.}{2001}]{bhw01}
{Baraffe} I.,  {Heger} A.,   {Woosley} S.~E.,  2001, \mn@doi [\apj]
  {10.1086/319808}, \href {http://adsabs.harvard.edu/abs/2001ApJ...550..890B}
  {550, 890}

\bibitem[\protect\citeauthoryear{{Barrow}, {Aykutalp}  \& {Wise}}{{Barrow}
  et~al.}{2018}]{bar18}
{Barrow} K. S.~S.,  {Aykutalp} A.,   {Wise} J.~H.,  2018, \mn@doi [Nature
  Astronomy] {10.1038/s41550-018-0569-y}, \href
  {https://ui.adsabs.harvard.edu/abs/2018NatAs...2..987B} {2, 987}

\bibitem[\protect\citeauthoryear{{Baumgarte} \& {Shapiro}}{{Baumgarte} \&
  {Shapiro}}{1999}]{baum99}
{Baumgarte} T.~W.,  {Shapiro} S.~L.,  1999, \mn@doi [\apj] {10.1086/308006},
  \href {http://adsabs.harvard.edu/abs/1999ApJ...526..941B} {526, 941}

\bibitem[\protect\citeauthoryear{{Bear} \& {Soker}}{{Bear} \&
  {Soker}}{2020}]{Soker2020}
{Bear} E.,  {Soker} N.,  2020, \mn@doi [\na] {10.1016/j.newast.2020.101438},
  \href {https://ui.adsabs.harvard.edu/abs/2020NewA...8101438B} {81, 101438}

\bibitem[\protect\citeauthoryear{{Begelman}, {Rossi}  \& {Armitage}}{{Begelman}
  et~al.}{2008}]{begel08}
{Begelman} M.~C.,  {Rossi} E.~M.,   {Armitage} P.~J.,  2008, \mn@doi [\mnras]
  {10.1111/j.1365-2966.2008.13344.x}, \href
  {http://adsabs.harvard.edu/abs/2008MNRAS.387.1649B} {387, 1649}

\bibitem[\protect\citeauthoryear{{Bromm} \& {Loeb}}{{Bromm} \&
  {Loeb}}{2003}]{bl03}
{Bromm} V.,  {Loeb} A.,  2003, \mn@doi [\apj] {10.1086/377529}, \href
  {http://adsabs.harvard.edu/abs/2003ApJ...596...34B} {596, 34}

\bibitem[\protect\citeauthoryear{{Butler}, {Lima}, {Baumgarte}  \&
  {Shapiro}}{{Butler} et~al.}{2018}]{but18}
{Butler} S.~P.,  {Lima} A.~R.,  {Baumgarte} T.~W.,   {Shapiro} S.~L.,  2018,
  \mn@doi [\mnras] {10.1093/mnras/sty834}, \href
  {http://adsabs.harvard.edu/abs/2018MNRAS.477.3694B} {477, 3694}

\bibitem[\protect\citeauthoryear{{Chandrasekhar}}{{Chandrasekhar}}{1964}]{chandra64}
{Chandrasekhar} S.,  1964, \mn@doi [\apj] {10.1086/147938}, \href
  {http://adsabs.harvard.edu/abs/1964ApJ...140..417C} {140, 417}

\bibitem[\protect\citeauthoryear{{Chen}, {Heger}, {Woosley}, {Almgren},
  {Whalen}  \& {Johnson}}{{Chen} et~al.}{2014}]{chen14b}
{Chen} K.-J.,  {Heger} A.,  {Woosley} S.,  {Almgren} A.,  {Whalen} D.~J.,
  {Johnson} J.~L.,  2014, \mn@doi [\apj] {10.1088/0004-637X/790/2/162}, \href
  {http://adsabs.harvard.edu/abs/2014ApJ...790..162C} {790, 162}

\bibitem[\protect\citeauthoryear{{Fan} et~al.,}{{Fan} et~al.}{2003}]{fan03}
{Fan} X.,  et~al., 2003, \mn@doi [\aj] {10.1086/368246}, \href
  {http://adsabs.harvard.edu/abs/2003AJ....125.1649F} {125, 1649}

\bibitem[\protect\citeauthoryear{{Fan} et~al.,}{{Fan} et~al.}{2006}]{fan06}
{Fan} X.,  et~al., 2006, \mn@doi [\aj] {10.1086/500296}, \href
  {http://adsabs.harvard.edu/abs/2006AJ....131.1203F} {131, 1203}

\bibitem[\protect\citeauthoryear{{Fowler}}{{Fowler}}{1964}]{fowler64}
{Fowler} W.~A.,  1964, \mn@doi [Reviews of Modern Physics]
  {10.1103/RevModPhys.36.545}, \href
  {http://adsabs.harvard.edu/abs/1964RvMP...36..545F} {36, 545}

\bibitem[\protect\citeauthoryear{{Fowler}}{{Fowler}}{1966}]{fowler66}
{Fowler} W.~A.,  1966, \mn@doi [\apj] {10.1086/148594}, \href
  {http://adsabs.harvard.edu/abs/1966ApJ...144..180F} {144, 180}

\bibitem[\protect\citeauthoryear{{Fuller}, {Woosley}  \& {Weaver}}{{Fuller}
  et~al.}{1986}]{fuller86}
{Fuller} G.~M.,  {Woosley} S.~E.,   {Weaver} T.~A.,  1986, \mn@doi [\apj]
  {10.1086/164452}, \href {http://adsabs.harvard.edu/abs/1986ApJ...307..675F}
  {307, 675}

\bibitem[\protect\citeauthoryear{{Haemmerl{\'e}}, {Woods}, {Klessen}, {Heger}
  \& {Whalen}}{{Haemmerl{\'e}} et~al.}{2018a}]{hle18b}
{Haemmerl{\'e}} L.,  {Woods} T.~E.,  {Klessen} R.~S.,  {Heger} A.,   {Whalen}
  D.~J.,  2018a, \mn@doi [\mnras] {10.1093/mnras/stx2919}, \href
  {http://adsabs.harvard.edu/abs/2018MNRAS.474.2757H} {474, 2757}

\bibitem[\protect\citeauthoryear{{Haemmerl{\'e}}, {Woods}, {Klessen}, {Heger}
  \& {Whalen}}{{Haemmerl{\'e}} et~al.}{2018b}]{hle18a}
{Haemmerl{\'e}} L.,  {Woods} T.~E.,  {Klessen} R.~S.,  {Heger} A.,   {Whalen}
  D.~J.,  2018b, \mn@doi [\apjl] {10.3847/2041-8213/aaa462}, \href
  {http://adsabs.harvard.edu/abs/2018ApJ...853L...3H} {853, L3}

\bibitem[\protect\citeauthoryear{{Hirano}, {Hosokawa}, {Yoshida}, {Umeda},
  {Omukai}, {Chiaki}  \& {Yorke}}{{Hirano} et~al.}{2014}]{hir13}
{Hirano} S.,  {Hosokawa} T.,  {Yoshida} N.,  {Umeda} H.,  {Omukai} K.,
  {Chiaki} G.,   {Yorke} H.~W.,  2014, \mn@doi [\apj]
  {10.1088/0004-637X/781/2/60}, \href
  {http://adsabs.harvard.edu/abs/2014ApJ...781...60H} {781, 60}

\bibitem[\protect\citeauthoryear{{Hirano}, {Hosokawa}, {Yoshida}, {Omukai}  \&
  {Yorke}}{{Hirano} et~al.}{2015}]{hir15}
{Hirano} S.,  {Hosokawa} T.,  {Yoshida} N.,  {Omukai} K.,   {Yorke} H.~W.,
  2015, \mn@doi [\mnras] {10.1093/mnras/stv044}, \href
  {http://adsabs.harvard.edu/abs/2015MNRAS.448..568H} {448, 568}

\bibitem[\protect\citeauthoryear{{Hirano}, {Hosokawa}, {Yoshida}  \&
  {Kuiper}}{{Hirano} et~al.}{2017}]{hir17}
{Hirano} S.,  {Hosokawa} T.,  {Yoshida} N.,   {Kuiper} R.,  2017, \mn@doi
  [Science] {10.1126/science.aai9119}, \href
  {http://adsabs.harvard.edu/abs/2017Sci...357.1375H} {357, 1375}

\bibitem[\protect\citeauthoryear{{Hosokawa}, {Yorke}, {Inayoshi}, {Omukai}  \&
  {Yoshida}}{{Hosokawa} et~al.}{2013}]{hos13}
{Hosokawa} T.,  {Yorke} H.~W.,  {Inayoshi} K.,  {Omukai} K.,   {Yoshida} N.,
  2013, \mn@doi [\apj] {10.1088/0004-637X/778/2/178}, \href
  {http://adsabs.harvard.edu/abs/2013ApJ...778..178H} {778, 178}

\bibitem[\protect\citeauthoryear{{Iben}}{{Iben}}{1963}]{iben63}
{Iben} Jr. I.,  1963, \mn@doi [\apj] {10.1086/147708}, \href
  {http://adsabs.harvard.edu/abs/1963ApJ...138.1090I} {138, 1090}

\bibitem[\protect\citeauthoryear{{Joggerst}, {Almgren}, {Bell}, {Heger},
  {Whalen}  \& {Woosley}}{{Joggerst} et~al.}{2010}]{jet09b}
{Joggerst} C.~C.,  {Almgren} A.,  {Bell} J.,  {Heger} A.,  {Whalen} D.,
  {Woosley} S.~E.,  2010, \mn@doi [\apj] {10.1088/0004-637X/709/1/11}, \href
  {http://adsabs.harvard.edu/abs/2010ApJ...709...11J} {709, 11}

\bibitem[\protect\citeauthoryear{{Johnson}, {Whalen}, {Fryer}  \&
  {Li}}{{Johnson} et~al.}{2012}]{jlj12a}
{Johnson} J.~L.,  {Whalen} D.~J.,  {Fryer} C.~L.,   {Li} H.,  2012, \mn@doi
  [\apj] {10.1088/0004-637X/750/1/66}, \href
  {http://adsabs.harvard.edu/abs/2012ApJ...750...66J} {750, 66}

\bibitem[\protect\citeauthoryear{{Johnson}, {Whalen}, {Li}  \&
  {Holz}}{{Johnson} et~al.}{2013a}]{jet13}
{Johnson} J.~L.,  {Whalen} D.~J.,  {Li} H.,   {Holz} D.~E.,  2013a, \mn@doi
  [\apj] {10.1088/0004-637X/771/2/116}, \href
  {http://adsabs.harvard.edu/abs/2013ApJ...771..116J} {771, 116}

\bibitem[\protect\citeauthoryear{{Johnson}, {Whalen}, {Even}, {Fryer}, {Heger},
  {Smidt}  \& {Chen}}{{Johnson} et~al.}{2013b}]{jet13a}
{Johnson} J.~L.,  {Whalen} D.~J.,  {Even} W.,  {Fryer} C.~L.,  {Heger} A.,
  {Smidt} J.,   {Chen} K.-J.,  2013b, \mn@doi [\apj]
  {10.1088/0004-637X/775/2/107}, \href
  {http://adsabs.harvard.edu/abs/2013ApJ...775..107J} {775, 107}

\bibitem[\protect\citeauthoryear{{Kitayama}, {Yoshida}, {Susa}  \&
  {Umemura}}{{Kitayama} et~al.}{2004}]{ket04}
{Kitayama} T.,  {Yoshida} N.,  {Susa} H.,   {Umemura} M.,  2004, \mn@doi [\apj]
  {10.1086/423313}, \href {http://adsabs.harvard.edu/abs/2004ApJ...613..631K}
  {613, 631}

\bibitem[\protect\citeauthoryear{{Latif} \& {Schleicher}}{{Latif} \&
  {Schleicher}}{2016}]{ls15}
{Latif} M.~A.,  {Schleicher} D.~R.~G.,  2016, \mn@doi [\aap]
  {10.1051/0004-6361/201527266}, \href
  {http://adsabs.harvard.edu/abs/2016A%26A...585A.151L} {585, A151}

\bibitem[\protect\citeauthoryear{{Latif} \& {Volonteri}}{{Latif} \&
  {Volonteri}}{2015}]{latif15b}
{Latif} M.~A.,  {Volonteri} M.,  2015, \mn@doi [\mnras]
  {10.1093/mnras/stv1337}, \href
  {http://adsabs.harvard.edu/abs/2015MNRAS.452.1026L} {452, 1026}

\bibitem[\protect\citeauthoryear{{Latif}, {Niemeyer}  \& {Schleicher}}{{Latif}
  et~al.}{2014a}]{lns14}
{Latif} M.~A.,  {Niemeyer} J.~C.,   {Schleicher} D.~R.~G.,  2014a, \mn@doi
  [\mnras] {10.1093/mnras/stu489}, \href
  {http://adsabs.harvard.edu/abs/2014MNRAS.440.2969L} {440, 2969}

\bibitem[\protect\citeauthoryear{{Latif}, {Bovino}, {Van Borm}, {Grassi},
  {Schleicher}  \& {Spaans}}{{Latif} et~al.}{2014b}]{latif14}
{Latif} M.~A.,  {Bovino} S.,  {Van Borm} C.,  {Grassi} T.,  {Schleicher}
  D.~R.~G.,   {Spaans} M.,  2014b, \mn@doi [\mnras] {10.1093/mnras/stu1230},
  \href {http://adsabs.harvard.edu/abs/2014MNRAS.443.1979L} {443, 1979}

\bibitem[\protect\citeauthoryear{{Latif}, {Khochfar}  \& {Whalen}}{{Latif}
  et~al.}{2020}]{latif20a}
{Latif} M.~A.,  {Khochfar} S.,   {Whalen} D.,  2020, \mn@doi [\apjl]
  {10.3847/2041-8213/ab7c61}, \href
  {https://ui.adsabs.harvard.edu/abs/2020ApJ...892L...4L} {892, L4}

\bibitem[\protect\citeauthoryear{{Latif}, {Khochfar}, {Schleicher}  \&
  {Whalen}}{{Latif} et~al.}{2021}]{latif21a}
{Latif} M.~A.,  {Khochfar} S.,  {Schleicher} D.,   {Whalen} D.~J.,  2021,
  \mn@doi [\mnras] {10.1093/mnras/stab2708}, \href
  {https://ui.adsabs.harvard.edu/abs/2021MNRAS.508.1756L} {508, 1756}

\bibitem[\protect\citeauthoryear{{Latif}, {Whalen}, {Khochfar}, {Herrington}
  \& {Woods}}{{Latif} et~al.}{2022a}]{latif22b}
{Latif} M.~A.,  {Whalen} D.~J.,  {Khochfar} S.,  {Herrington} N.~P.,   {Woods}
  T.~E.,  2022a, \mn@doi [\nat] {10.1038/s41586-022-04813-y}, \href
  {https://ui.adsabs.harvard.edu/abs/2022Natur.607...48L} {607, 48}

\bibitem[\protect\citeauthoryear{{Latif}, {Whalen}  \& {Khochfar}}{{Latif}
  et~al.}{2022b}]{latif22a}
{Latif} M.~A.,  {Whalen} D.,   {Khochfar} S.,  2022b, \mn@doi [\apj]
  {10.3847/1538-4357/ac3916}, \href
  {https://ui.adsabs.harvard.edu/abs/2022ApJ...925...28L} {925, 28}

\bibitem[\protect\citeauthoryear{{Lodato} \& {Natarajan}}{{Lodato} \&
  {Natarajan}}{2006}]{ln06}
{Lodato} G.,  {Natarajan} P.,  2006, \mn@doi [\mnras]
  {10.1111/j.1365-2966.2006.10801.x}, \href
  {http://adsabs.harvard.edu/abs/2006MNRAS.371.1813L} {371, 1813}

\bibitem[\protect\citeauthoryear{{Luo}, {Ardaneh}, {Shlosman}, {Nagamine},
  {Wise}  \& {Begelman}}{{Luo} et~al.}{2018}]{luo18}
{Luo} Y.,  {Ardaneh} K.,  {Shlosman} I.,  {Nagamine} K.,  {Wise} J.~H.,
  {Begelman} M.~C.,  2018, \mn@doi [\mnras] {10.1093/mnras/sty362}, \href
  {http://adsabs.harvard.edu/abs/2018MNRAS.476.3523L} {476, 3523}

\bibitem[\protect\citeauthoryear{{Matsuoka} et~al.,}{{Matsuoka}
  et~al.}{2019}]{mats19}
{Matsuoka} Y.,  et~al., 2019, \mn@doi [\apjl] {10.3847/2041-8213/ab0216}, \href
  {https://ui.adsabs.harvard.edu/abs/2019ApJ...872L...2M} {872, L2}

\bibitem[\protect\citeauthoryear{{Montero}, {Janka}  \& {M{\"u}ller}}{{Montero}
  et~al.}{2012}]{montero12}
{Montero} P.~J.,  {Janka} H.-T.,   {M{\"u}ller} E.,  2012, \mn@doi [\apj]
  {10.1088/0004-637X/749/1/37}, \href
  {http://adsabs.harvard.edu/abs/2012ApJ...749...37M} {749, 37}

\bibitem[\protect\citeauthoryear{{Moriya}, {Chen}, {Nakajima}, {Tominaga}  \&
  {Blinnikov}}{{Moriya} et~al.}{2021}]{mor21}
{Moriya} T.~J.,  {Chen} K.-J.,  {Nakajima} K.,  {Tominaga} N.,   {Blinnikov}
  S.~I.,  2021, \mn@doi [\mnras] {10.1093/mnras/stab622}, \href
  {https://ui.adsabs.harvard.edu/abs/2021MNRAS.503.1206M} {503, 1206}

\bibitem[\protect\citeauthoryear{{Mortlock} et~al.,}{{Mortlock}
  et~al.}{2011}]{mort11}
{Mortlock} D.~J.,  et~al., 2011, \mn@doi [\nat] {10.1038/nature10159}, \href
  {http://adsabs.harvard.edu/abs/2011Natur.474..616M} {474, 616}

\bibitem[\protect\citeauthoryear{{Mu{\~n}oz}, {Takhistov}, {Witte}  \&
  {Fuller}}{{Mu{\~n}oz} et~al.}{2021}]{mun21}
{Mu{\~n}oz} V.,  {Takhistov} V.,  {Witte} S.~J.,   {Fuller} G.~M.,  2021,
  \mn@doi [\jcap] {10.1088/1475-7516/2021/11/020}, \href
  {https://ui.adsabs.harvard.edu/abs/2021JCAP...11..020M} {2021, 020}

\bibitem[\protect\citeauthoryear{{Nagele}, {Umeda}, {Takahashi}, {Yoshida}  \&
  {Sumiyoshi}}{{Nagele} et~al.}{2021}]{nag21}
{Nagele} C.,  {Umeda} H.,  {Takahashi} K.,  {Yoshida} T.,   {Sumiyoshi} K.,
  2021, \mn@doi [\mnras] {10.1093/mnras/stab2592}, \href
  {https://ui.adsabs.harvard.edu/abs/2021MNRAS.508..828N} {508, 828}

\bibitem[\protect\citeauthoryear{{Nagele}, {Umeda}, {Takahashi}, {Yoshida}  \&
  {Sumiyoshi}}{{Nagele} et~al.}{2022}]{nag22}
{Nagele} C.,  {Umeda} H.,  {Takahashi} K.,  {Yoshida} T.,   {Sumiyoshi} K.,
  2022, \mn@doi [\mnras] {10.1093/mnras/stac2495}, \href
  {https://ui.adsabs.harvard.edu/abs/2022MNRAS.tmp.2306N} {}

\bibitem[\protect\citeauthoryear{{Natarajan}, {Pacucci}, {Ferrara}, {Agarwal},
  {Ricarte}, {Zackrisson}  \& {Cappelluti}}{{Natarajan} et~al.}{2017}]{nat17}
{Natarajan} P.,  {Pacucci} F.,  {Ferrara} A.,  {Agarwal} B.,  {Ricarte} A.,
  {Zackrisson} E.,   {Cappelluti} N.,  2017, \mn@doi [\apj]
  {10.3847/1538-4357/aa6330}, \href
  {http://adsabs.harvard.edu/abs/2017ApJ...838..117N} {838, 117}

\bibitem[\protect\citeauthoryear{{Pacucci}, {Ferrara}, {Volonteri}  \&
  {Dubus}}{{Pacucci} et~al.}{2015}]{pac15}
{Pacucci} F.,  {Ferrara} A.,  {Volonteri} M.,   {Dubus} G.,  2015, \mn@doi
  [\mnras] {10.1093/mnras/stv2196}, \href
  {http://adsabs.harvard.edu/abs/2015MNRAS.454.3771P} {454, 3771}

\bibitem[\protect\citeauthoryear{{Patrick}, {Whalen}, {Elford}  \&
  {Latif}}{{Patrick} et~al.}{2020}]{pat21a}
{Patrick} S.~J.,  {Whalen} D.~J.,  {Elford} J.~S.,   {Latif} M.~A.,  2020,
  arXiv e-prints, \href {https://ui.adsabs.harvard.edu/abs/2020arXiv201211612P}
  {p. arXiv:2012.11612}

\bibitem[\protect\citeauthoryear{{Patrick}, {Whalen}, {Elford}  \&
  {Latif}}{{Patrick} et~al.}{2021}]{pat21b}
{Patrick} S.,  {Whalen} D.~J.,  {Elford} J.~S.,   {Latif} M.,  2021, \mnras, in
  prep

\bibitem[\protect\citeauthoryear{{Paxton}, {Bildsten}, {Dotter}, {Herwig},
  {Lesaffre}  \& {Timmes}}{{Paxton} et~al.}{2011}]{paxt11}
{Paxton} B.,  {Bildsten} L.,  {Dotter} A.,  {Herwig} F.,  {Lesaffre} P.,
  {Timmes} F.,  2011, \mn@doi [\apjs] {10.1088/0067-0049/192/1/3}, \href
  {http://adsabs.harvard.edu/abs/2011ApJS..192....3P} {192, 3}

\bibitem[\protect\citeauthoryear{{Paxton} et~al.,}{{Paxton}
  et~al.}{2013}]{paxt13}
{Paxton} B.,  et~al., 2013, \mn@doi [\apjs] {10.1088/0067-0049/208/1/4}, \href
  {http://adsabs.harvard.edu/abs/2013ApJS..208....4P} {208, 4}

\bibitem[\protect\citeauthoryear{{Paxton} et~al.,}{{Paxton}
  et~al.}{2015}]{Paxton15}
{Paxton} B.,  et~al., 2015, \mn@doi [\apjs] {10.1088/0067-0049/220/1/15}, \href
  {https://ui.adsabs.harvard.edu/abs/2015ApJS..220...15P} {220, 15}

\bibitem[\protect\citeauthoryear{{Paxton} et~al.,}{{Paxton}
  et~al.}{2018}]{paxt18}
{Paxton} B.,  et~al., 2018, \mn@doi [\apjs] {10.3847/1538-4365/aaa5a8}, \href
  {https://ui.adsabs.harvard.edu/abs/2018ApJS..234...34P} {234, 34}

\bibitem[\protect\citeauthoryear{{Paxton} et~al.,}{{Paxton}
  et~al.}{2019}]{Paxton19}
{Paxton} B.,  et~al., 2019, \mn@doi [\apjs] {10.3847/1538-4365/ab2241}, \href
  {https://ui.adsabs.harvard.edu/abs/2019ApJS..243...10P} {243, 10}

\bibitem[\protect\citeauthoryear{Potekhin \& Chabrier}{Potekhin \&
  Chabrier}{2010}]{potekhin10}
Potekhin A.~Y.,  Chabrier G.,  2010, Contributions to Plasma Physics, 50, 82

\bibitem[\protect\citeauthoryear{{Regan} \& {Haehnelt}}{{Regan} \&
  {Haehnelt}}{2009}]{rh09b}
{Regan} J.~A.,  {Haehnelt} M.~G.,  2009, \mn@doi [\mnras]
  {10.1111/j.1365-2966.2009.14579.x}, \href
  {http://adsabs.harvard.edu/abs/2009MNRAS.396..343R} {396, 343}

\bibitem[\protect\citeauthoryear{{Regan}, {Wise}, {Woods}, {Downes}, {O'Shea}
  \& {Norman}}{{Regan} et~al.}{2020}]{ret20}
{Regan} J.~A.,  {Wise} J.~H.,  {Woods} T.~E.,  {Downes} T.~P.,  {O'Shea} B.~W.,
    {Norman} M.~L.,  2020, arXiv:2008.08090, \href
  {https://ui.adsabs.harvard.edu/abs/2020arXiv200808090R} {p. arXiv:2008.08090}

\bibitem[\protect\citeauthoryear{Rogers \& Nayfonov}{Rogers \&
  Nayfonov}{2002}]{rogers02}
Rogers F.,  Nayfonov A.,  2002, The Astrophysical Journal, 576, 1064

\bibitem[\protect\citeauthoryear{{Sakurai}, {Hosokawa}, {Yoshida}  \&
  {Yorke}}{{Sakurai} et~al.}{2015}]{sak15}
{Sakurai} Y.,  {Hosokawa} T.,  {Yoshida} N.,   {Yorke} H.~W.,  2015, \mn@doi
  [\mnras] {10.1093/mnras/stv1346}, \href
  {http://adsabs.harvard.edu/abs/2015MNRAS.452..755S} {452, 755}

\bibitem[\protect\citeauthoryear{{Sakurai}, {Vorobyov}, {Hosokawa}, {Yoshida},
  {Omukai}  \& {Yorke}}{{Sakurai} et~al.}{2016}]{sak16b}
{Sakurai} Y.,  {Vorobyov} E.~I.,  {Hosokawa} T.,  {Yoshida} N.,  {Omukai} K.,
  {Yorke} H.~W.,  2016, \mn@doi [\mnras] {10.1093/mnras/stw637}, \href
  {http://adsabs.harvard.edu/abs/2016MNRAS.459.1137S} {459, 1137}

\bibitem[\protect\citeauthoryear{Saumon, Chabrier  \& van Horn}{Saumon
  et~al.}{1995}]{saumon1995}
Saumon D.,  Chabrier G.,   van Horn H.~M.,  1995, The astrophysical journal
  supplement series, 99, 713

\bibitem[\protect\citeauthoryear{{Schauer}, {Whalen}, {Glover}  \&
  {Klessen}}{{Schauer} et~al.}{2015}]{anna15}
{Schauer} A.~T.~P.,  {Whalen} D.~J.,  {Glover} S.~C.~O.,   {Klessen} R.~S.,
  2015, \mn@doi [\mnras] {10.1093/mnras/stv2117}, \href
  {http://adsabs.harvard.edu/abs/2015MNRAS.454.2441S} {454, 2441}

\bibitem[\protect\citeauthoryear{{Schauer} et~al.,}{{Schauer}
  et~al.}{2017a}]{anna17}
{Schauer} A.~T.~P.,  et~al., 2017a, \mn@doi [\mnras] {10.1093/mnras/stx264},
  \href {http://adsabs.harvard.edu/abs/2017MNRAS.467.2288S} {467, 2288}

\bibitem[\protect\citeauthoryear{{Schauer}, {Regan}, {Glover}  \&
  {Klessen}}{{Schauer} et~al.}{2017b}]{srg17}
{Schauer} A.~T.~P.,  {Regan} J.,  {Glover} S.~C.~O.,   {Klessen} R.~S.,  2017b,
  \mn@doi [\mnras] {10.1093/mnras/stx1915}, \href
  {http://adsabs.harvard.edu/abs/2017MNRAS.471.4878S} {471, 4878}

\bibitem[\protect\citeauthoryear{{Schober}, {Schleicher}, {Federrath},
  {Glover}, {Klessen}  \& {Banerjee}}{{Schober} et~al.}{2012}]{schob12}
{Schober} J.,  {Schleicher} D.,  {Federrath} C.,  {Glover} S.,  {Klessen}
  R.~S.,   {Banerjee} R.,  2012, \mn@doi [\apj] {10.1088/0004-637X/754/2/99},
  \href {http://adsabs.harvard.edu/abs/2012ApJ...754...99S} {754, 99}

\bibitem[\protect\citeauthoryear{{Shapiro} \& {Teukolsky}}{{Shapiro} \&
  {Teukolsky}}{1979}]{st79}
{Shapiro} S.~L.,  {Teukolsky} S.~A.,  1979, \mn@doi [\apjl] {10.1086/183134},
  \href {http://adsabs.harvard.edu/abs/1979ApJ...234L.177S} {234, L177}

\bibitem[\protect\citeauthoryear{{Shi}, {Fuller}  \& {Halzen}}{{Shi}
  et~al.}{1998}]{sfh98}
{Shi} X.,  {Fuller} G.~M.,   {Halzen} F.,  1998, \mn@doi [Physical Review
  Letters] {10.1103/PhysRevLett.81.5722}, \href
  {http://adsabs.harvard.edu/abs/1998PhRvL..81.5722S} {81, 5722}

\bibitem[\protect\citeauthoryear{{Smidt}, {Whalen}, {Johnson}, {Surace}  \&
  {Li}}{{Smidt} et~al.}{2018}]{smidt18}
{Smidt} J.,  {Whalen} D.~J.,  {Johnson} J.~L.,  {Surace} M.,   {Li} H.,  2018,
  \mn@doi [\apj] {10.3847/1538-4357/aad7b8}, \href
  {http://adsabs.harvard.edu/abs/2018ApJ...865..126S} {865, 126}

\bibitem[\protect\citeauthoryear{{Smith}, {Becerra}, {Bromm}  \&
  {Hernquist}}{{Smith} et~al.}{2017}]{aaron17}
{Smith} A.,  {Becerra} F.,  {Bromm} V.,   {Hernquist} L.,  2017, \mn@doi
  [\mnras] {10.1093/mnras/stx1993}, \href
  {http://adsabs.harvard.edu/abs/2017MNRAS.472..205S} {472, 205}

\bibitem[\protect\citeauthoryear{{Smith}, {Regan}, {Downes}, {Norman}, {O'Shea}
   \& {Wise}}{{Smith} et~al.}{2018}]{srd18}
{Smith} B.~D.,  {Regan} J.~A.,  {Downes} T.~P.,  {Norman} M.~L.,  {O'Shea}
  B.~W.,   {Wise} J.~H.,  2018, \mn@doi [\mnras] {10.1093/mnras/sty2103}, \href
  {http://adsabs.harvard.edu/abs/2018MNRAS.480.3762S} {480, 3762}

\bibitem[\protect\citeauthoryear{{Sugimura}, {Matsumoto}, {Hosokawa}, {Hirano}
  \& {Omukai}}{{Sugimura} et~al.}{2020}]{sug20}
{Sugimura} K.,  {Matsumoto} T.,  {Hosokawa} T.,  {Hirano} S.,   {Omukai} K.,
  2020, \mn@doi [\apjl] {10.3847/2041-8213/ab7d37}, \href
  {https://ui.adsabs.harvard.edu/abs/2020ApJ...892L..14S} {892, L14}

\bibitem[\protect\citeauthoryear{{Sun}, {Ruiz}  \& {Shapiro}}{{Sun}
  et~al.}{2018}]{sun18}
{Sun} L.,  {Ruiz} M.,   {Shapiro} S.~L.,  2018, \mn@doi [\prd]
  {10.1103/PhysRevD.98.103008}, \href
  {http://adsabs.harvard.edu/abs/2018PhRvD..98j3008S} {98, 103008}

\bibitem[\protect\citeauthoryear{{Surace} et~al.,}{{Surace}
  et~al.}{2018}]{sur18a}
{Surace} M.,  et~al., 2018, \mn@doi [\apjl] {10.3847/2041-8213/aaf80d}, \href
  {https://ui.adsabs.harvard.edu/abs/2018ApJ...869L..39S} {869, L39}

\bibitem[\protect\citeauthoryear{{Surace}, {Zackrisson}, {Whalen}, {Hartwig},
  {Glover}, {Woods}, {Heger}  \& {Glover}}{{Surace} et~al.}{2019}]{sur19a}
{Surace} M.,  {Zackrisson} E.,  {Whalen} D.~J.,  {Hartwig} T.,  {Glover}
  S.~C.~O.,  {Woods} T.~E.,  {Heger} A.,   {Glover} S.~C.~O.,  2019, \mn@doi
  [\mnras] {10.1093/mnras/stz1956}, \href
  {https://ui.adsabs.harvard.edu/abs/2019MNRAS.488.3995S} {488, 3995}

\bibitem[\protect\citeauthoryear{{Susa}}{{Susa}}{2013}]{susa13}
{Susa} H.,  2013, \mn@doi [\apj] {10.1088/0004-637X/773/2/185}, \href
  {http://adsabs.harvard.edu/abs/2013ApJ...773..185S} {773, 185}

\bibitem[\protect\citeauthoryear{{Susa}, {Umemura}  \& {Hasegawa}}{{Susa}
  et~al.}{2009}]{suh09}
{Susa} H.,  {Umemura} M.,   {Hasegawa} K.,  2009, \mn@doi [\apj]
  {10.1088/0004-637X/702/1/480}, \href
  {http://adsabs.harvard.edu/abs/2009ApJ...702..480S} {702, 480}

\bibitem[\protect\citeauthoryear{{The Lynx Team}}{{The Lynx Team}}{2018}]{lynx}
{The Lynx Team} 2018, arXiv e-prints, \href
  {https://ui.adsabs.harvard.edu/abs/2018arXiv180909642T} {p. arXiv:1809.09642}

\bibitem[\protect\citeauthoryear{{Timmes} \& {Swesty}}{{Timmes} \&
  {Swesty}}{2000}]{ts00}
{Timmes} F.~X.,  {Swesty} F.~D.,  2000, \mn@doi [\apjs] {10.1086/313304}, \href
  {http://adsabs.harvard.edu/abs/2000ApJS..126..501T} {126, 501}

\bibitem[\protect\citeauthoryear{Townsend}{Townsend}{2019}]{mesasdk}
Townsend R.,  2019, \mn@doi [] {10.5281/zenodo.2669543}

\bibitem[\protect\citeauthoryear{{Umeda}, {Hosokawa}, {Omukai}  \&
  {Yoshida}}{{Umeda} et~al.}{2016}]{um16}
{Umeda} H.,  {Hosokawa} T.,  {Omukai} K.,   {Yoshida} N.,  2016, \mn@doi
  [\apjl] {10.3847/2041-8205/830/2/L34}, \href
  {http://adsabs.harvard.edu/abs/2016ApJ...830L..34U} {830, L34}

\bibitem[\protect\citeauthoryear{{Vikaeus}, {Whalen}  \&
  {Zackrisson}}{{Vikaeus} et~al.}{2022}]{vik22a}
{Vikaeus} A.,  {Whalen} D.~J.,   {Zackrisson} E.,  2022, \mn@doi [\apjl]
  {10.3847/2041-8213/ac7802}, \href
  {https://ui.adsabs.harvard.edu/abs/2022ApJ...933L...8V} {933, L8}

\bibitem[\protect\citeauthoryear{{Vink}, {de Koter}  \& {Lamers}}{{Vink}
  et~al.}{2001}]{vink01}
{Vink} J.~S.,  {de Koter} A.,   {Lamers} H.~J.~G.~L.~M.,  2001, \mn@doi [\aap]
  {10.1051/0004-6361:20010127}, \href
  {http://adsabs.harvard.edu/abs/2001A%26A...369..574V} {369, 574}

\bibitem[\protect\citeauthoryear{{Volonteri} \& {Begelman}}{{Volonteri} \&
  {Begelman}}{2010}]{vb10}
{Volonteri} M.,  {Begelman} M.~C.,  2010, \mn@doi [\mnras]
  {10.1111/j.1365-2966.2010.17359.x}, \href
  {http://adsabs.harvard.edu/abs/2010MNRAS.409.1022V} {409, 1022}

\bibitem[\protect\citeauthoryear{{Wang} et~al.,}{{Wang} et~al.}{2021}]{wang21}
{Wang} F.,  et~al., 2021, \mn@doi [\apjl] {10.3847/2041-8213/abd8c6}, \href
  {https://ui.adsabs.harvard.edu/abs/2021ApJ...907L...1W} {907, L1}

\bibitem[\protect\citeauthoryear{{Whalen} \& {Fryer}}{{Whalen} \&
  {Fryer}}{2012}]{wf12}
{Whalen} D.~J.,  {Fryer} C.~L.,  2012, \mn@doi [\apjl]
  {10.1088/2041-8205/756/1/L19}, \href
  {http://adsabs.harvard.edu/abs/2012ApJ...756L..19W} {756, L19}

\bibitem[\protect\citeauthoryear{{Whalen} \& {Mezcua}}{{Whalen} \&
  {Mezcua}}{2022}]{wet22b}
{Whalen} D.~J.,  {Mezcua} M.,  2022, \mnras, in prep

\bibitem[\protect\citeauthoryear{{Whalen}, {Abel}  \& {Norman}}{{Whalen}
  et~al.}{2004}]{wan04}
{Whalen} D.,  {Abel} T.,   {Norman} M.~L.,  2004, \mn@doi [\apj]
  {10.1086/421548}, \href {http://adsabs.harvard.edu/abs/2004ApJ...610...14W}
  {610, 14}

\bibitem[\protect\citeauthoryear{{Whalen}, {Fryer}, {Holz}, {Heger}, {Woosley},
  {Stiavelli}, {Even}  \& {Frey}}{{Whalen} et~al.}{2013a}]{wet12a}
{Whalen} D.~J.,  {Fryer} C.~L.,  {Holz} D.~E.,  {Heger} A.,  {Woosley} S.~E.,
  {Stiavelli} M.,  {Even} W.,   {Frey} L.~H.,  2013a, \mn@doi [\apjl]
  {10.1088/2041-8205/762/1/L6}, \href
  {http://adsabs.harvard.edu/abs/2013ApJ...762L...6W} {762, L6}

\bibitem[\protect\citeauthoryear{{Whalen}, {Johnson}, {Smidt}, {Heger}, {Even}
  \& {Fryer}}{{Whalen} et~al.}{2013b}]{wet13b}
{Whalen} D.~J.,  {Johnson} J.~L.,  {Smidt} J.,  {Heger} A.,  {Even} W.,
  {Fryer} C.~L.,  2013b, \mn@doi [\apj] {10.1088/0004-637X/777/2/99}, \href
  {http://adsabs.harvard.edu/abs/2013ApJ...777...99W} {777, 99}

\bibitem[\protect\citeauthoryear{{Whalen}, {Smidt}, {Even}, {Woosley}, {Heger},
  {Stiavelli}  \& {Fryer}}{{Whalen} et~al.}{2014}]{wet13d}
{Whalen} D.~J.,  {Smidt} J.,  {Even} W.,  {Woosley} S.~E.,  {Heger} A.,
  {Stiavelli} M.,   {Fryer} C.~L.,  2014, \mn@doi [\apj]
  {10.1088/0004-637X/781/2/106}, \href
  {http://adsabs.harvard.edu/abs/2014ApJ...781..106W} {781, 106}

\bibitem[\protect\citeauthoryear{{Whalen}, {Mezcua}, {Meiksin}, {Hartwig}  \&
  {Latif}}{{Whalen} et~al.}{2020a}]{wet20a}
{Whalen} D.~J.,  {Mezcua} M.,  {Meiksin} A.,  {Hartwig} T.,   {Latif} M.~A.,
  2020a, \mn@doi [\apjl] {10.3847/2041-8213/ab9a30}, \href
  {https://ui.adsabs.harvard.edu/abs/2020ApJ...896L..45W} {896, L45}

\bibitem[\protect\citeauthoryear{{Whalen}, {Surace}, {Bernhardt}, {Zackrisson},
  {Pacucci}, {Ziegler}  \& {Hirschmann}}{{Whalen} et~al.}{2020b}]{wet20b}
{Whalen} D.~J.,  {Surace} M.,  {Bernhardt} C.,  {Zackrisson} E.,  {Pacucci} F.,
   {Ziegler} B.,   {Hirschmann} M.,  2020b, \mn@doi [\apjl]
  {10.3847/2041-8213/ab9d29}, \href
  {https://ui.adsabs.harvard.edu/abs/2020ApJ...897L..16W} {897, L16}

\bibitem[\protect\citeauthoryear{{Whalen}, {Mezcua}, {Patrick}, {Meiksin}  \&
  {Latif}}{{Whalen} et~al.}{2021}]{wet21a}
{Whalen} D.~J.,  {Mezcua} M.,  {Patrick} S.~J.,  {Meiksin} A.,   {Latif} M.~A.,
   2021, \mn@doi [\apjl] {10.3847/2041-8213/ac35e6}, \href
  {https://ui.adsabs.harvard.edu/abs/2021ApJ...922L..39W} {922, L39}

\bibitem[\protect\citeauthoryear{{Wise}, {Turk}  \& {Abel}}{{Wise}
  et~al.}{2008}]{wta08}
{Wise} J.~H.,  {Turk} M.~J.,   {Abel} T.,  2008, \mn@doi [\apj]
  {10.1086/588209}, \href {http://adsabs.harvard.edu/abs/2008ApJ...682..745W}
  {682, 745}

\bibitem[\protect\citeauthoryear{{Wise}, {Regan}, {O'Shea}, {Norman}, {Downes}
  \& {Xu}}{{Wise} et~al.}{2019}]{wise19}
{Wise} J.~H.,  {Regan} J.~A.,  {O'Shea} B.~W.,  {Norman} M.~L.,  {Downes}
  T.~P.,   {Xu} H.,  2019, \mn@doi [\nat] {10.1038/s41586-019-0873-4}, \href
  {http://adsabs.harvard.edu/abs/2019Natur.566...85W} {566, 85}

\bibitem[\protect\citeauthoryear{Wolf \& Schwab}{Wolf \& Schwab}{2017}]{pymesa}
Wolf B.,  Schwab J.,  2017, \mn@doi [] {10.5281/zenodo.826958}

\bibitem[\protect\citeauthoryear{{Woods}, {Heger}, {Whalen}, {Haemmerl{\'e}}
  \& {Klessen}}{{Woods} et~al.}{2017}]{tyr17}
{Woods} T.~E.,  {Heger} A.,  {Whalen} D.~J.,  {Haemmerl{\'e}} L.,   {Klessen}
  R.~S.,  2017, \mn@doi [\apjl] {10.3847/2041-8213/aa7412}, \href
  {http://adsabs.harvard.edu/abs/2017ApJ...842L...6W} {842, L6}

\bibitem[\protect\citeauthoryear{{Woods} et~al.,}{{Woods}
  et~al.}{2019}]{titans}
{Woods} T.~E.,  et~al., 2019, \mn@doi [Publications of the Astronomical Society
  of Australia] {10.1017/pasa.2019.14}, \href
  {https://ui.adsabs.harvard.edu/abs/2019PASA...36...27W} {36, e027}

\bibitem[\protect\citeauthoryear{{Woods}, {Heger}  \& {Haemmerl{\'e}}}{{Woods}
  et~al.}{2020}]{tyr20a}
{Woods} T.~E.,  {Heger} A.,   {Haemmerl{\'e}} L.,  2020, \mn@doi [\mnras]
  {10.1093/mnras/staa763}, \href
  {https://ui.adsabs.harvard.edu/abs/2020MNRAS.494.2236W} {494, 2236}

\bibitem[\protect\citeauthoryear{{Woods}, {Patrick}, {Whalen}  \&
  {Heger}}{{Woods} et~al.}{2021a}]{tyr22a}
{Woods} T.~E.,  {Patrick} S.,  {Whalen} D.~J.,   {Heger} A.,  2021a, arXiv
  e-prints, \href {https://ui.adsabs.harvard.edu/abs/2021arXiv211209142W} {p.
  arXiv:2112.09142}

\bibitem[\protect\citeauthoryear{{Woods}, {Patrick}, {Elford}, {Whalen}  \&
  {Heger}}{{Woods} et~al.}{2021b}]{tyr21a}
{Woods} T.~E.,  {Patrick} S.,  {Elford} J.~S.,  {Whalen} D.~J.,   {Heger} A.,
  2021b, \mn@doi [\apj] {10.3847/1538-4357/abfaf9}, \href
  {https://ui.adsabs.harvard.edu/abs/2021ApJ...915..110W} {915, 110}

\bibitem[\protect\citeauthoryear{{Wu} et~al.,}{{Wu} et~al.}{2015}]{wu15}
{Wu} X.-B.,  et~al., 2015, \mn@doi [\nat] {10.1038/nature14241}, \href
  {http://adsabs.harvard.edu/abs/2015Natur.518..512W} {518, 512}

\bibitem[\protect\citeauthoryear{{Zhu}, {Li}, {Li}, {Maji}, {Yajima},
  {Schneider}  \& {Hernquist}}{{Zhu} et~al.}{2020}]{zhu20}
{Zhu} Q.,  {Li} Y.,  {Li} Y.,  {Maji} M.,  {Yajima} H.,  {Schneider} R.,
  {Hernquist} L.,  2020, arXiv e-prints, \href
  {https://ui.adsabs.harvard.edu/abs/2020arXiv201201458Z} {p. arXiv:2012.01458}

\makeatother
\end{thebibliography}

%%%%%%%%%%%%%%%%% APPENDICES %%%%%%%%%%%%%%%%%%%%%

% \appendix

% \section{Some extra material}

% If you want to present additional material which would interrupt the flow of the main paper,
% it can be placed in an Appendix which appears after the list of references.

%%%%%%%%%%%%%%%%%%%%%%%%%%%%%%%%%%%%%%%%%%%%%%%%%%

% Don't change these lines
\bsp	% typesetting comment
\label{lastpage}

\end{document}